\documentclass[1p,10pt]{elsarticle}

\usepackage{amsmath,amssymb,bm}

\journal{Annals of Physics}

\newcommand{\av}[1]{\ensuremath{\langle#1\rangle}}
\newcommand{\Av}[1]{\ensuremath{\big<#1\big>}}
\newcommand{\AV}[1]{\ensuremath{\Big<#1\Big>}}
\newcommand{\bra}[1]{\ensuremath{\bm{\langle}#1\bm{|}}}
\newcommand{\ket}[1]{\ensuremath{\bm{|}#1\bm{\rangle}}}

\begin{document}

\begin{frontmatter}

\title{The plasmon--polariton scattering by random-impedance surface defects:
       The interplay between localization and outflow\tnoteref{t1}}

\tnotetext[t1]{Yu.~T. and N.~K. gratefully acknowledge support from the National Research Foundation of Ukraine, Project No. 2020.02/0149 ``Quantum phenomena in the interaction of electromagnetic waves with solid-state nanostructures''.}

\author{Yu.~V. Tarasov\corref{cor1}}
\ead{yuriy.tarasov@gmail.com}

\author{O.~M. Stadnyk}

\author{N.~Kvitka}

\cortext[cor1]{Corresponding author}

\affiliation{organization={O.~Ya.~Usikov Institute for Radiophysics and Electronics of the National Academy of Sciences of Ukraine},
addressline={12~Acad.~Proskura~Str.},
postcode={61085},
city={Kharkiv},
country={Ukraine}}
\begin{abstract}
We study the scattering of TM-polarized surface plasmon--polariton (SPP) by the finite section of flat metal-vacuum interface with a random impedance. We analyze the solution to the integral equation that connect the scattered field and the incident plasmon--polariton, and is valid for any strength of the scattering and dissipative characteristics of the conducting half-space.
We show that the norm of the intermode scattering operator as a measure of scattering strength is not only determined by the parameters of the random impedance (the variance, correlation radius, the length of the heterogeneous section of the interface), but  also crucially depends on the metal conductivity. For a small norm of the integral operator, the incident surface plasmon polariton radiates effectively into vacuum, resulting in excitation of quasi-isotropic Norton-type waves above the conducting surface. The intensity of the leaking field is expressed in terms of the pair correlation function of the impedance, whose dependence on wave numbers of incident and scattered waves demonstrates the possibility to observe a phenomenon similar to Wood's anomalies of wave scattering by periodic gratings. Under strong scattering regime, the radiation into the upper half-space is highly suppressed and the SPP wave is mainly backscattered from the heterogeneous surface segment. For the lossless conducting half-space, the surface plasmon--polariton becomes unstable for arbitrarily small fluctuations of the conductor polarizability. The mirroring should also take place at small norm of the scattering operator, yet in this case it is related to Anderson's localization of the SPP within the disordered section.
\end{abstract}
%

\begin{keyword}
plasmon--polariton\sep random impedance\sep scattering\sep Anderson localization\sep radiation leakage

\PACS 68.65.-k \sep 73.20.Fz \sep 78.67.-n
\end{keyword}

\end{frontmatter}

\section{Introduction}
With the development of nanotechnology, modern optics in recent decades has been replenished with a new promising branch --- the optics of surface electromagnetic waves (plasmonics).
These waves exist only in TM-polarization and are associated with collective oscillations of free electrons in noble metals. The emergence of plasmonics was primarily due to the discovery of Wood's anomalies ~\cite{Wood1902} in metal diffraction gratings, as well as Fano resonances \cite{Fano41}. Significant progress in understanding the properties of SPP has been achieved in the works of Hessel and Oliner \cite{Hessel65}, and now plasmonics is an extensive field of research (see, for example, Refs.~\cite{Maier07,Zayats05,Han13,SaridChallener2010,EnochBonod12,bib:Shahbaz13,Vinogradov2020} and references therein).

Interest in plasmon polaritons is due to their unique properties associated with high spatial localization and the possibility to strongly enhance the propagating field. These waves can be effectively excited by light and supported by a metal surface. Great potential for miniaturizing various optical devices has led to the need to study the SPP propagation not only on plane surfaces but also on surfaces of other geometries, such as metal strips and waveguides, ribs on plane surfaces, wires on plane and non-plane single and multiple interfaces, and many others.

SPP scattering on inhomogeneous surfaces is also associated with the the existence of their localized states ~\cite{MarFreiSimLesk01}, similar to the localized states in the one-dimensional version of Anderson's theory \cite{Anderson58}. Importance of this effect is due to the potential application of SPPs for ultra dense recording of information and ultra fast its processing by optical devices, where delocalized modes could serve for carrying the energy, while localized modes would provide the possibility to concentrate and store the energy in small regions of the system ~\cite {StockmanFaleevBergman01}; hence SPPs can transfer information faster and with less energy loss than occurs in conventional electrical circuits.

The study of surface excitation scattering on imperfect surfaces \cite{Maradudin1986} is needed to solve a number of scientific and technical issues during the development of plasmonics. Starting with Mandelshtamm's 1913 work on light scattering by the surfaces of liquids ~\cite {Mandelshtam13}, scattering mechanisms for the SPPs was studied mainly in the approximation of \emph{single} scattering events. For rough surfaces, this approximation uses either a perturbation theory which assumes small deviations of the surface relief from a flat one, or the Kirchhoff approximation, in which the surface curvature serves as a small parameter (results on gradient scattering that does not use the Kirchhoff approximation see in Refs.~\cite{bib:GorTarShost13, TarShost15}). It is worth noting that in \cite{Kretschmann72} polarization and angular dependences of the scattered light intensity were considered in the approximation when the rough surface is replaced by a smooth one and an additional layer of dipole currents and SPP are excited. The same dependences were observed thereafter in the experiment \cite{SimonGuha76}.

Since the discovery of inverse coherent light reflection associated with so-called weak localization the researchers' interest has shifted to the \emph{multiple}-scattering processes, and a~number of effects were discovered such as the reflection suppression, the transformation of polarization, the appearance of forbidden Bragg bands in the SPP spectrum. Besides SPP in simple conductor configurations (comparison of approaches, as well as their combinations and methods of calculation in this case can be found in the works ~\cite{Nikitin07,PolancoFitzMaradudin13}), multiple wave scattering by fluctuations in surface relief and/or metal impedance was corroborated experimentally \cite{Horstmann77} and has become one of the most interesting and intriguing problem. Very interesting effects due to multiple scattering of SPP by surface nanoparticles are of great importance for the sensitivity of plasmonic sensing techniques \cite{SonderBozhev2003, SonderBozhev2004, Foreman19, BerkForeman21, BerkForemanPRB21}.

The method of Green's function, impedance boundary condition, the method of reduced Rayleigh equation, ``exact'' numerical integration, and other methods for SPP calculations have gained a~widespread popularity.

The complex nature of the SPP propagation has led to the need to introduce simple parameters that would make it relatively easy to model, calculate and design electromagnetic (EM) devices. Among the reliable parameters, the impedance is one of the most commonly used characteristic, since it significantly simplifies the problem solution and gives the possibility to avoid solving the electrodynamic equations directly in the metal. Moreover, the impedance description is proven to be applicable in problems of wave scattering at curved surfaces, both regular and random \cite{OngCelliMarv94}. A brief overview of the development of the impedance concept can be found in Ref.~\cite{Kaiser2013}.

Regardless of the applied method --- the accurate consideration of the surface relief or its modeling by the impedance boundary condition --- the solution of the specific problem results in the need to solve integral equation describing the SPP scattering. In particular, such an equation arises when using the impedance boundary condition after substituting the solution of the Helmholtz equation in the form of decomposition by natural (in the simplest case, plane) waves and the transition to momentum representation, see Eq.~\eqref{R(q)-mainEq} below or Ref.~\cite{Depine92}.

The result of the analysis of the integral equations strongly depends on the metal model used in solving such problems. In Ref.~\cite{EnochBonod12} (subsection 1.2.5), the discrepancy between the theoretical (obtained in the \textit{ideally conducting metals} model) and experimentally observed results on the diffraction of $p$-polarized light by metal lattices is reported, while for $s$-polarized light no such discrepancy was noticed.
Among possible reasons for the discrepancy there was pointed out, in particular, the discrepancy between the model of an ideal metal and a real metal with finite conductivity.

In Ref.~\cite{MarFreiSimLesk01}, the possibility of strong localization of SPPs propagating along metal surface with random finite conductivity was considered. The goal was to create a surface that would suppress the transformation of the SPPs into a bulk EM wave, though later, in paper~\cite{Coello08} the result was questioned.

\begin{figure}[h!!]
	\centering
	\scalebox{.7}[.7]{\includegraphics{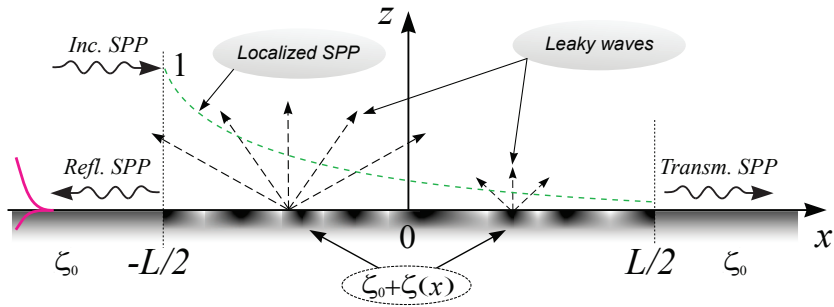}}
	\caption{Geometry of the problem of SPP scattering by a plane boundary segment of finite length $L$ with randomly modulated surface
		impedance.
		\hfill\label{fig1}}
\end{figure}

In the present work, we study the plasmon--polariton scattering by a metal strip of finite width $L$ with random impedance inside it, see Fig.~\ref{fig1}. The impedance approximation is used due to its universality in taking into account the fluctuations of both the electric parameters of the metal and the relief of the metal-vacuum interface \cite{OngCelliMarv94}, whereas the strip geometry is chosen to allow for the scattering to be maximally reduced to one dimension. The wave vector of the plasmon lies in the metal plane and is directed perpendicular to the scattering region of the interface. Radiation of the surface wave into free space (the ``leakage''), caused by fluctuations in the surface impedance, is considered on the background of absorption related to the finite conductivity of the metal.
The issues related to the SPP radiation leakage are important both for understanding the principles of operation and for finding the ways for increasing the functionality of the so-called leakage radiation microscopes \cite{Merlo2014, Hohenau2011}, in particular, in such interesting areas as microscopy at the diffraction limit \cite{Takayama2017} and the observation of non-transparent samples~\cite{GRANDIDIER2010}.

To analyze the problem, we use the integral equation for the scattered field which is expressed in terms of the Fourier components of the EM field on the surface. The main difficulty in solving the governing equation is due to the presence of an integral term which is usually considered as a small perturbation. We suggest and substantiate a specific criterion for estimating the scattering intensity for the SPP, which is the norm of the operator mixing the intermediate scattering states. We calculate the scattering pattern and show that for weak mixing of surface and bulk scattered modes the radiated energy is proportional to the Fourier transform of the two-point correlation function of the impedance, the argument of which is equal to the difference between the SPP wave number and the projection of the leaking harmonics wave vector directed to the observation point. Such a dependence has obvious similarities with the angular dependence of the field scattered by periodic reflecting gratings, where Wood ~\cite{Wood1902} discovered the diffraction anomalies in the form of peaks of radiation intensity.

The work is organized as follows. In section \ref{Formulation} we state the problem, choose the model of finding the Helmholtz equation solution in the form of unperturbed and scattered fields, and formulate boundary conditions (BCs) for the field scattered by an inhomogeneous area of the surface. Section \ref{Model_solution} is devoted to the choice and justification of a trial solution, which is used to obtain the expressions for the full field of the scattered waves. In section \ref {Full_Sol}, we obtain the solution to the equation for leaking field in the operator form, which is expressed in terms of the field at the metal-vacuum interface, and introduce correlation characteristics of the random complex impedance. In short section~\ref{Asymptota}, we give the asymptotics of the scattering amplitude for the cases of weak and strong coupling of the surface and volume parts of the scattered field, which are analyzed in the next section~\ref{ScattDiagr}, where the most important results are summarized. In this section, we obtain the scattering patterns of the surface plasmon--polariton in the limit of weak coupling of the scattered components and predict the phenomenon of totally specular reflection of the SPP from a surface area even with weakly fluctuating impedance. The conclusion is devoted to the discussion of the results. In Appendices, we provide some details of the calculation of the mode-mixing operator norm and of SPP dynamic scattering lengths.
\section{Formulation of the problem}
\label{Formulation}
Consider a two-dimensional problem of scattering of SPP wave excited by a line source
(for example, of slot nature~\cite{bib:Tejeira05, bib:Tejeira07}) on impedance boundary between metal and vacuum. In the finite region of the boundary the impedance is assumed to depend on one coordinate only and to vary randomly. It can be represented by the function consisting of two complex-valued terms,
\begin{equation}\label{Z}
	Z_s(x)=\zeta_0+\zeta(x)\ .
\end{equation}
The first term, $\zeta_0$, is constant everywhere on $x$-axis while the second term, $\zeta(x)$, is a random function of $x$ with a nonzero value only in the interval ${\mathbb{L}:\,x\in[-L/2,L/2]}$ and is continuous at its ends. The average value of $\zeta(x)$ in this interval is chosen to be zero, $\av{\zeta(x)}=0$.

Surface plasmon--polariton is a $p$-polarized (TM) wave the only nonzero magnetic field component of which, $H_y$, satisfies standard Helmholtz equation and impedance boundary condition \cite{bib:Leontovich85}. By denoting $H_y\equiv H (\mathbf{r})$, where ${\mathbf{r}=(x,z)}$ is the two-dimensional radius-vector, we state the problem of determining the magnetic field over a conducting half-space in the form of equation
\begin{subequations}\label{initial_eq_Psi}
	\begin{equation}\label{Helmholtz_eq}
		\left(\Delta+k^2\right)H(\mathbf{r})=0\ ,
	\end{equation}
	where $\Delta$ denotes (two-dimensional) Laplace operator, $k=\omega/c$, and the boundary condition of the following form,
	\begin{equation}\label{Impedance_BC}
		\left(\frac{\partial H}{\partial z}+ik\big[\zeta_0+\zeta(x)\big]H\right)
		\Bigg|_{z=0}=0\ .
	\end{equation}
\end{subequations}
We assume function $\zeta(x)$ to be complex-valued and define its two-point correlation properties using a couple of equalities given below, see Eqs.~\eqref{Basic_corrs(x)}.

In addition, we must also set the conditions on the $x$-axis, which depend on the physical statement of the problem. We will solve the \emph{scattering problem} for the surface plasmon--polariton incident on an inhomogeneous segment $\mathbb{L}$ of the boundary from the left and scattered by this region in the form of plane harmonics in all directions. At large distances from the scattering segment, Sommerfeld's radiation conditions must be fulfilled~\cite{Schot92, Atkinson49}, which imply that the scattered field contains outgoing harmonics only. For the total field, which includes also the field of the oncoming wave, the conditions are formulated taking account of both the nature and the location of the SPP real source.

For a coordinate-independent surface impedance with $\mathrm{Im}\,\zeta_0<0$, the problem~\eqref{initial_eq_Psi} supports two solutions of the following form,
\begin{align}\label{RightLeft_SPP}
	& H_{spp}^{(\pm)}(\mathbf{r})= \mathcal{A}^{(\pm)}
	\exp\big(\pm ik_{spp}x-i\zeta_0 kz\big)\ ,\\
	\notag
	& k_{spp}= k\sqrt{1-\zeta_0^2}\ .
\end{align}
Since we are interested in SPPs propagating \emph{quasi-freely} along the dielectric-metal boundary at distances much larger than their wavelength, from \eqref{RightLeft_SPP} it follows that the following inequality must be satisfied,
\begin{subequations}\label{SPP}
	\begin{equation}\label{freeSPP}
		|k''_{spp}| \ll k'_{spp}\ .
	\end{equation}
For good metals this happens at not too high frequencies, when the impedance modulus is small as compared to unity.
The imaginary part of the impedance of metals is negative and does not exceed unity in absolute value (see, for example, Ref.~\cite{Palik98}). Assuming the limitation \eqref{freeSPP}, which corresponds to weak dissipation in the metal, i.\,e., to inequality
	\begin{equation}\label{Low_dissip}
		\zeta'_0\ll |\zeta''_0|\ ,
	\end{equation}
\end{subequations}
by expanding quantity $k_{spp}$ in Eq.~\eqref{RightLeft_SPP} we obtain the following asymptotic expression for the SPP propagation constant,
\begin{equation}\label{SPP_asymp_wavenumber}
	k_{spp} = k'_{spp}
	+ ik''_{spp}\approx k\left(\sqrt{1+{(\zeta_0'')}^2}
	+ i\frac{\zeta_0'|\zeta_0''|}{\sqrt{1+{\zeta_0''}^2}}\right)\ .
\end{equation}
Based on this formula, the SPP damping length along the metal-dielectric interface, which is associated with \textit{dissipation} in the metal half-space, can be introduced, namely,
\begin{equation}\label{L(SPP)_dis}
	L^{(spp)}_{dis}= \big|k''_{spp}\big|^{-1}
	= \frac{\sqrt{1+{\zeta_0''}^2}}{k\zeta_0'|\zeta_0''|}\ .
\end{equation}
Provided that the conditions for the existence of quasi-free SPP in the case of \emph{uniform} impedance ($\zeta(x)\equiv 0$) are satisfied, we will seek the solution of problem~\eqref{initial_eq_Psi} as a sum of conditionally unperturbed \textit{trial} solution  $H_0(\mathbf{r})$ with plasmon--polariton field structure and the ``perturbation'' $h(\mathbf{r})$, which is expected to be not a purely surface wave,
\begin{align}\label{h-BC}
	H(\mathbf{r})=H_0(\mathbf{r})+h(\mathbf{r})\ .
\end{align}
Since it comes to solution of SPP type, we choose function $H_0(\mathbf{r})$ to have the form conventional for \textit{one-dimensional} scattering problems. Specifically, in the region $x<-L/2$ we present $H_0(\mathbf{r})\equiv H_0^{(l)}(\mathbf{r})$ as a sum of the SPP with unit (at $z=0$) amplitude, which is incident onto the disordered segment of the boundary from the left, and the reflected plasmon--polariton whose propagation and \emph{dissipative} damping are in the opposite direction,
\begin{subequations}\label{H_0^lrc}
\begin{align}\label{H_0^<}
  H_0^{(l)}(\mathbf{r})=\Big\{\exp\left[ik_{spp}\big(x+ L/2\big)\right]+
  r_-\exp\left[-ik_{spp}\big(x+ L/2\big)\right]\Big\}
  \exp\left[-i\zeta_0 kz\right]\ .
\end{align}
To the right of the region $\mathbb{L}$ we choose $H_0(\mathbf{r})\equiv H_0^{(r)}(\mathbf{r})$ in the form of \emph{transmitted} SPP,
\begin{align}\label{H_0^>}
  H_0^{(r)}(\mathbf{r})=t_+\exp\left[ik_{spp}\big(x- L/2\big)-i\zeta_0 kz\right]\ .
\end{align}
In the intermediate region  $x\in\mathbb{L}$, we model trial field $H_0(\mathbf{r})\equiv H_0^{(int)}(\mathbf{r})$ as a~superposition of counterpropagating SPPs with modulated amplitudes to be determined, which flow in opposite directions of the $x$-axis,
\begin{align}\label{H_0^in}
  & H_0^{(int)}(\mathbf{r})=\mathcal{H}^{(int)}_0(x)\exp\big(-i\zeta_0 kz\big)\ .
\end{align}
\end{subequations}
Here we can choose function $\mathcal{H}^{(int)}_0(x)$ in any convenient form, since it is suggested as not a \textit{true} solution of the problem~\eqref{initial_eq_Psi} within interval $\mathbb{L}$ but only as a part of this solution. Moreover, we will seek function $H_0^{(int)}(\mathbf{r})$ in the form \eqref{H_0^in} only close enough to the metal surface, at height coordinate $z\lesssim 1/k|\zeta_0|$. At larger vertical distances we can regard $H_0^{(int)}(\mathbf{r})$ to be quite arbitrary, yet decreasing with the increase in $z$ faster than in power form. The missing part of the complete solution, which is not confined exponentially near the metal surface, is assumed to be included in function~$h(\mathbf{r})$.

The choice of a trial solution in the form \eqref{H_0^lrc} is motivated, on the one hand, by the desire to have as a zero approximation the SPP which is scattered by random fluctuations of electrodynamic properties of the boundary \textit{along one coordinate} only, in our case along axis $x$, without radiation into the free (upper) half-space. One more reason for this choice is to describe the fluctuation addition to the impedance in the boundary conditions \eqref{Impedance_BC} in terms of some \emph{effective} one-dimensional potentials included in the master equation. In case of successful implementation of this program, we would be able to correctly describe the effect of Anderson localization of the polariton, the physical mechanism of which is associated with the interference of \textit{surface} harmonics of the SPP's field during their multiple backward rescattering due to one-dimensional impedance inhomogeneities, without emission into the free space.

Indeed, the localization of Anderson type is widely known to be described as a result of interference of plane harmonics undergoing multiple back-rescattering by inhomogeneities of a one-dimensional medium (see, for example, Refs.~\cite{Berez74,AbrikRyzh78}). For SPPs specifically, some aspects of such a localization were studied in Ref.~\cite{FreiYurk93}, where the authors have explained by this phenomenon a peak in the backscattering of the plane wave incident on a~surface with random impedance from the free half-space. It is essential, however, that in \cite{FreiYurk93} the \emph{bulk} wave scattering from the entirely infinite plane with random surface impedance was studied while in the present work we consider a \textit{surface} plasma wave scattering from the bounded random-impedance area. In such a system, no deterministic simple functions but the plane waves can be chosen as a zero approximation for the scattering problem.  The latter functions, although they belong to a different mathematical class than square-integrable functions, also are localized due to the scattering by one-dimensional random inhomogeneities, and moreover, they remain random at any point of the propagation axis \cite{LifGredPast88}.

In what follows we assume that at $z>0$ the model trial field \eqref{H_0^lrc} taken as a~zero approximation for leaking field $h(\mathbf{r})$ satisfies Helmholtz equation \eqref{Helmholtz_eq}. The boundary condition (BC) for this field is the condition \eqref{Impedance_BC}, in which exact function  $H(\mathbf{r})$ should be replaced with trial function  $H_0(\mathbf{r})$ with removing the random part of the surface impedance. With such a choice of a ``zero approximation'' for the total field, the field $h(\mathbf{r})$ must also satisfy equation \eqref{Helmholtz_eq}. Outside the interval~$\mathbb{L}$ on ~$x$-axis it satisfies the BC \emph{\`{a}~la}~\eqref{Impedance_BC} with $\zeta(x)\equiv 0$, whereas within this interval we obtain the boundary condition for $h(\mathbf{r})$ in the form
\begin{align}\label{psi(x)-BC_in_L}
	\left(\frac{\partial h(\mathbf{r})}{\partial z}+ik\big[\zeta_0+\zeta(x)\big]h(\mathbf{r})\right)
	\Bigg|_{z=0}=-ik\zeta(x)\mathcal{H}_0^{(int)}(x) \qquad\ (x\in\mathbb{L})\ ,
\end{align}
which connects the component emitted into the upper half-space and, as its source, the seed field ~\eqref{H_0^in}.

Since the solutions of SPP structure are already accounted for in formulas \eqref{H_0^<} and \eqref{H_0^>}, at $|x|>L/2$ the leaking field on the surface can be thought absent, $h(x,0)\equiv 0$. As to the conditions this field satisfies at $z>0$, due to the finite nature of interval $\mathbb{L}$, which, according to ~\eqref{psi(x)-BC_in_L}, can be considered as a source for the leaking field at large distances ${|\mathbf{r}|=\sqrt{x^2+z^2}\gg L}$, we will assume function $h(\mathbf{r})$ to satisfy the radiation conditions.
%
\section{The trial field within the disordered boundary segment}
\label{Model_solution}
\subsection{The basic idea for choosing the trial field form}
\label{Basic_idea}
As regards the specific choice of function $\mathcal{H}^{(int)}_0(x)$ in formula \eqref{H_0^in}, it would be desirable that it include information respecting Anderson localization of the polariton penetrated into disordered segment $\mathbb{L}$ through its left border ${x=-L/2}$. This implies the effect of \textit{random component} of the surface impedance to be somehow included into $\mathcal{H}^{(int)}_0(x)$. If we were originally looking for solution to equation~\eqref{Helmholtz_eq} with the dependence on $z$ described by function $\exp\big\{-i[\zeta_0+\zeta(x)]kz\big\}$, we would satisfy BC \eqref{Impedance_BC} at once on the entire $x$-axis. However, for the solution in region $x\in\mathbb{L}$ we would get an equation whose solution at $z>0$ could not be obtained in analytical form. In such a way, we would not be able to detect the leaking harmonics which are actually present in practice. So, we deliberately refuse to satisfy the correct BC in region $\mathbb{L}$ automatically at every point in order to obtain a closed form of the solution by choosing a simple dependence on $z$ of trial function $H_0(\mathbf{r})$ on the entire $x$-axis yet adding the term $h(\mathbf{r})$ which, on the one hand, should ensure the fulfillment of the correct BC in the entire region $\mathbb{L}$, and, on the other hand, would not have a purely surface nature but would include the harmonics propagating into free space to infinity in the direction of~$z$-axis.

To model factor $\mathcal{H}_0^{(int)}(x)$ in Eq.~\eqref{H_0^in} reasonably well consider first the \textit{hypothetical} problem of a wave with local (in $x$) plasmon--polariton field structure, which satisfy simultaneously equation
\eqref{Helmholtz_eq} and BC \eqref{Impedance_BC}. We will calculate the field of this wave right at the metal surface ($z\to 0$), without taking into account the leaking component $h(\mathbf{r})$. The field of such a \textit{speculative} (\textit{s}) wave will be sought in the form of a~function whose dependence on $z$ has (locally in $x$ coordinate) strictly SPP form, viz.,
\begin{align}\label{H_0^in-mod}
	&  H^{(s)}(\mathbf{r})=\mathcal{H}^{(s)}(x)
	\exp\Big\{-i\big[\zeta_0+\zeta(x)\big] kz\Big\}\ .
\end{align}
Certainly, this field cannot be considered as a true one for $\forall z>0$. It includes, in addition to the surface-nature trial solution~\eqref{H_0^in},  some part of the ``radiation'' field $h(\mathbf{r})$. The latter is generated by the near-surface field \textit{not locally} in $x$, because in our approach each point of the surface in segment $\mathbb{L}$ can be considered as an independent emitter.

By substituting function \eqref{H_0^in-mod} into equation \eqref{Helmholtz_eq} and setting thereafter ${z=0}$ (to remove from the wave equation the ``inconvenient'' terms proportional to $z$ and $z^2$), we obtain the following equation for $\mathcal{H}^{(s)}(x)$:
\begin{subequations}\label{Potentials_V1V2}
	\begin{equation}\label{H_0(x)-eq}
		\left\{\frac{\partial^2 }{\partial x^2}+
		\left[\varkappa^2_{s}-V_1(x)-V_2(x)\right]\right\}\mathcal{H}^{(s)}(x)=0\ .
	\end{equation}
	Here, the notations are introduced
\begin{align}
\label{varkappa_spp}
 & \varkappa^2_{s} = k^2_{spp}-
	k^2\,\bm{\Xi}^2\ ,\qquad \bm{\Xi}^2=\Av{\zeta^2(x)}\ ,\\
 \label{V1(x)}
 & V_1(x) =2k^2\zeta_0\zeta(x)\ ,\\	
	\label{V2(x)}
 & V_2(x) =k^2\left[\zeta^2(x)-\bm{\Xi}^2\right]\ .
\end{align}
\end{subequations}
Equation \eqref{H_0(x)-eq} is similar in structure to Schr\"odinger equation in one dimension, in which the quantity $\varkappa^2_{s}$ ($\mathrm{Im}\,\varkappa^2_{s}>0$) plays the role of (generally speaking, complex) ``energy'', while the terms $V_{1,2}(x)$ stand for the generalized complex-valued potentials the action of which results in the \textit{scattering} (and thus in modification) of the ``unperturbed'' solution stemming from Eq.~\eqref{H_0(x)-eq} in the absence of these potentials. Both potentials in Eq.~\eqref{H_0(x)-eq} are intentionally constructed so as to have zero mean values. Just for random potentials of such a property the solution to 1D stochastic equations can be obtained to any order of perturbation theory, which is necessary to exactly account for eigenstate localization in 1D starting from the only pre-known regular extended wave functions \cite{Berez74}.

Taking account of complex nature of wave parameter $\varkappa_{s}$, which is associated with the presence of complex terms  $k^2_{spp}$ and $\bm{\Xi}^2$ in formula ~\eqref{varkappa_spp}, the solution to Eq.~\eqref{H_0(x)-eq} unperturbed by the above-mentioned potentials can be presented as a superposition of \textit{right-moving} and \textit{left-moving} harmonics $\exp(\pm i\varkappa'_{s}x)$ (${\varkappa'_{s}=\mathrm{Re}\,\varkappa_{s}>0}$). When solving the problems of one-dimensional wave transport in limited area contacting with two perfect half-lines we will set the BC for each of the contra-moving harmonics at only one of the interval ends (no matter at which one), thereby ensuring the fulfillment of the causality principle. The solution at the other end of the region can be found from controlling equation~\eqref{H_0(x)-eq}.

In the case of real-valued random potentials, equation \eqref{H_0(x)-eq} in an infinitely long one-dimensional system admits exponentially localized solutions only \cite{GoldMolchPstur77}. The system we are considering is finite (segment~$\mathbb{L}$) and open, which implies that the controlling wave operator is non-Hermitian. We cannot rely on the \textit{strict} discreteness of such system spectrum, which would indicate the localized nature of the eigenfunctions at any degree of potential complexity including the case where the referred to potentials are completely real. However, even in open one-dimensional systems the localization (of interference nature) still manifests itself, in particular, through exponential dependence of the conductance (transmission coefficient) on the system size (see, for example, Refs.~\cite{bib:MakTar98,bib:MakTar01}), in contrast to the power dependence typical for diffusive wave transport.

To reveal Anderson localization of eigenstates of a 1D system is mathematically rather difficult task since it requires summation of a large number of terms in the perturbation theory series, even if the scattering, from physical point of view, can be regarded as weak. The terms in the series (Feynman diagrams) which describe \textit{multiple} backward scattering of the initially free states, contribute almost equally to the observed quantities~\cite{Berez74}. As a result of the interference of multiply backscattered extended harmonics, exponentially localized states with everywhere dense yet discrete spectrum are formed in infinite 1D disordered systems \cite{GoldMolchPstur77}.

Let us try to find function $\mathcal{H}^{(s)}(x)$,  which we will use to model the factor $\mathcal{H}^{(int)}_0(x)$ from the Eq. \eqref{H_0^in}, as a sum of weakly modulated quasi-free harmonics  $\exp(\pm i\varkappa'_{s} x)$, which propagate along $x$-axis in opposite directions,

\begin{equation}\label{H^in->pi_gamma}
	\mathcal{H}^{(s)}(x)\equiv \mathcal{H}^{(int)}_0(x)=\pi(x)e^{i\varkappa'_{s} x}-
	i\gamma(x)e^{-i\varkappa'_{s} x}\ .
\end{equation}
By the term ``weakly modulated'' we mean that functions $\pi(x)$ and $\gamma(x)$ are smooth as compared to the next-standing exponentials, which is the essence of weak scattering (WS) approximation. With this conjecture, after inserting \eqref{H^in->pi_gamma} into equation \eqref{H_0(x)-eq} and neglecting small second derivatives of smooth ``amplitudes'' $\pi(x)$ and $\gamma(x)$, we obtain the following set of first-order equations for these functions, viz.,
\begin{subequations}\label{pi_gamma-eqs-new}
	\begin{align}
		\label{pi_gamma-eq1-new}
		& \frac{d\pi(x)}{dx}+\frac{\varkappa^2_{s}-{\varkappa'}^2_{s}}{2i\varkappa'_{s}}\pi(x)+
		i\eta(x)\pi(x)+\xi_-(x)\gamma(x)=0\ ,\\
		\label{pi_gamma-eq2-new}
		& \frac{d\gamma(x)}{dx}-\frac{\varkappa^2_{s}-{\varkappa'}^2_{s}}{2i\varkappa'_{s}}
		\gamma(x)-i\eta(x)\gamma(x)+\xi_+(x)\pi(x)=0\ .
	\end{align}
\end{subequations}
In \eqref{pi_gamma-eqs-new}, we have introduced the following spatially smoothed (partially integrated) random functions,
\begin{subequations}\label{eta_xi-def-new}
	\begin{align}
		\label{eta-def-new}
		\eta(x) & =\frac{1}{2 \varkappa^{\,'}_{s}}\int\limits_{x-l}^{x+l} \frac{d x'}{2 l} \big[V_1(x')+V_2(x')\big]\ ,\\
		\label{xi-def-new}
		\xi_{\pm}(x) & =\frac{1}{2 \varkappa^{\,'}_{s}}\int\limits_{x-l}^{x+l} \frac{d x'}{2 l}
		e^{\pm 2i\varkappa'_{s} x'}\big[V_1(x')+V_2(x')\big]\ .
	\end{align}
\end{subequations}
Equations \eqref{pi_gamma-eqs-new} with functions \eqref{eta_xi-def-new} are valid provided the set of inequalities is satisfied
\begin{equation}\label{lambda<l<L_sc-new}
	{\varkappa\,'_{s}}^{-1}\ll l\ll L^{(sc)},L^{(dis)}\ \ ,
\end{equation}
meaning that scattering length $L^{(sc)}$ (the characteristic spatial variation scale of smooth factors $\pi(x)$ and $\gamma(x)$) as well as the dissipation length of the polariton due to the in-metal absorption, $L^{(dis)}={\varkappa''_{s}}^{-1}\approx L^{(spp)}_{dis}$, substantially exceed wavelength ${\varkappa\,'_{s}}^{-1}$. The scattering for which the conditions \eqref{lambda<l<L_sc-new} are met in common is conventionally thought of as a weak one.

Equations \eqref{pi_gamma-eqs-new} can be simplified if we introduce, besides functions $\pi(x)$ and $\gamma(x)$, new modulating functions renormalized in the following way,
\begin{equation}\label{Tilde_pi_gamma}
	\begin{cases}
		\begin{aligned}
			\widetilde{\pi}(x) &= e^{\varkappa''_{s}x}\pi(x)\ ,\\
			\widetilde{\gamma}(x) &= e^{-\varkappa''_{s}x}\gamma(x)\ .
		\end{aligned}
	\end{cases}
\end{equation}
For these functions another set of equations follows, viz.,
\begin{subequations}\label{pi_gamma-eqs-new2}
	\begin{align}
		\label{pi_gamma-eq1-new2}
		& \frac{d\widetilde{\pi}(x)}{dx}+ i\eta(x)\widetilde{\pi}(x)+
		\widetilde{\xi}_-(x)\widetilde{\gamma}(x)=0\ ,\\
		\label{pi_gamma-eq2-new2}
		& \frac{d\widetilde{\gamma}(x)}{dx}-i\eta(x)\widetilde{\gamma}(x)+
		\widetilde{\xi}_+(x)\widetilde{\pi}(x)=0\ ,
	\end{align}
\end{subequations}
where random functions $\widetilde{\xi}_{\pm}(x)$ are related to the primordial functions \eqref{xi-def-new} by equalities
\begin{equation}\label{Tilde_xi->xi}
	\widetilde{\xi}_{\pm}(x)=e^{\mp 2\varkappa''_{s}x}\xi_{\pm}(x)\ .
\end{equation}
%
\subsection{Boundary conditions for functions $\widetilde{\pi}(x)$ and $\widetilde{\gamma}(x)$}
\label{Tilde_pi_gamma-BCs}
%
Dynamic equations \eqref{pi_gamma-eqs-new2} can be resolved in the general functional form if we set boundary conditions at the ends of the interval $\mathbb{L}$. The natural boundary condition on the assumed vertical boundaries passing through points ${x=\pm L/2}$ is the continuity of tangential components of the magnetic and electric fields. The sought-for field $H(\mathbf{r})$ is defined as the $y$-component of the \textit{magnetic} field of a TM-polarized wave, and the derivative of this component with respect to the coordinate $x$ is proportional to $z$-component of the \textit{electric} field. The latter component is tangential at the boundary between the regular and the disordered regions on the metal surface, and therefore the conditions for matching the magnetic and electric fields at points  $\mathbf{r}=(\pm L/2,0)$ can be written as the following set of equalities,
\begin{subequations}\label{Field_Matchig_conds-n}
	\begin{align}
		\label{HyEz->right-n}
		\begin{aligned}
			\begin{cases}
				& \mathcal{H}^{(int)}_0(L/2)+h(L/2,0)=t_+\ ,\\
				& {\mathcal{H}^{(int)}_0}'(L/2)+h'_x(L/2,0)=ik_{spp}t_+\ ,
			\end{cases}
		\end{aligned}
	\end{align}
	\begin{align}
		\label{HyEz->left-n}
		\qquad\qquad\
		\begin{aligned}
			\begin{cases}
				& \mathcal{H}^{(int)}_0(-L/2)+h(-L/2,0)=1+r_-\ ,\\
				& {\mathcal{H}^{(int)}_0}'(-L/2)+h'_x(-L/2,0)=ik_{spp}(1-r_-)\ .
			\end{cases}
		\end{aligned}
	\end{align}
\end{subequations}
From BC \eqref{psi(x)-BC_in_L}, given the continuity of function $\zeta(x)$ at the end points of $\mathbb{L}$, it can be shown that $h'_x(\pm L/2,0)=h(\pm L/2,0)= 0$. Equations \eqref{Field_Matchig_conds-n} are thus reduced to the following ones,
\begin{subequations}\label{Field_Matchig-2n}
	\begin{align}
		\label{HyEz->right-2n}
		\begin{aligned}
			\begin{cases}
				& \mathcal{H}^{(int)}_0(L/2)=t_+\ ,\\[3pt]
                & d\mathcal{H}^{(int)}_0/dx\Big|_{x=L/2}=ik_{spp}t_+\ ,
			\end{cases}
		\end{aligned}
	\end{align}
	\begin{align}
		\label{HyEz->left-2n}
		\qquad\qquad\
		\begin{aligned}
			\begin{cases}
				& \mathcal{H}^{(int)}_0(-L/2)=1+r_-\ ,\\[3pt]
				& d\mathcal{H}^{(int)}_0/dx\Big|_{x=-L/2}=ik_{spp}(1-r_-)\ .
			\end{cases}
		\end{aligned}
	\end{align}
\end{subequations}
By joining  the values of function \eqref{H^in->pi_gamma} and its derivative with the values and the derivatives of functions \eqref{H_0^<} and \eqref{H_0^>} at the end points of segment~$\mathbb{L}$, we obtain the following relationships at its ``plus'' end,
\begin{subequations}\label{tilpi,tilgamma-BCs+}
	\begin{align}
		\label{tilpi-BC+}
		& \widetilde{\pi}(L/2)=\frac{t_+}{2}
		  \left(\frac{k_{spp}}{\varkappa'_{s}}+1\right) e^{-i\varkappa_{s}L/2}\ ,\\
		\label{tilgamma-BC+}
		& \widetilde{\gamma}(L/2)=\frac{t_+}{2i}
		\left(\frac{k_{spp}}{\varkappa'_{s}}-1\right) e^{i\varkappa_{s}L/2}\ .
\end{align}
\end{subequations}
Boundary conditions on the left, ``minus''  boundary of segment $\mathbb{L}$ are obtained by a similar fitting the tangential components of magnetic and electric fields:
\begin{subequations}\label{tilpi,tilgamma,G-BC-}
	\begin{align}
		\label{pi-BC-}
		\pi(-L/2) &= \frac{1}{2}\left(\frac{k_{spp}} {\varkappa'_{s}}+1\right) \Big(1-r_- \mathcal{R}_s\Big) e^{i\varkappa'_{s}L/2}\ ,\\
		\label{g-BC-}
		\gamma(-L/2) &= \frac{1}{2i}\left(\frac{k_{spp}} {\varkappa'_{s}}-1\right) \Big(1-r_-/\mathcal{R}_s\Big) e^{-i\varkappa'_{s}L/2}\ .
	\end{align}
\end{subequations}
Here, to make formulas less cumbersome, we use the notation
\begin{align}\label{Euscr_R+}
	\mathcal{R}_s=\frac{k_{spp}-\varkappa\,'_{s}}
	{k_{spp}+\varkappa\,'_{s}}\ .
\end{align}
This quantity is definitely nonzero due to the mismatch of phase velocities of the wave in contacting domains (in the theory of waveguide systems it is associated with the so-called \textit{index mismatch}). In quantum mechanics, the quantity \eqref{Euscr_R+} is called the reflection coefficient from a potential step with particle energies $k^2_{spp}$ and $\varkappa'^2_{s}$ on opposite sides of it \cite{bib:LandauLifshitz77}.
Boundary values \eqref{tilpi,tilgamma,G-BC-} can be also found just by solving equations \eqref{pi_gamma-eqs-new2} with boundary conditions set at the ``plus'' end only.
%
\subsection{Correlation properties of random functions $\eta(x)$ and $\widetilde{\xi}_\pm(x)$}
\label{Corr_props}
%
Since the solution of our problem, being a functional of random impedance, is also of random nature, it is feasible to obtain meaningful information about it by applying at certain stage the procedure of statistical averaging. The necessary step for this is the setting up correlation properties of the primary random functions (fields) on which this solution is functionally dependent. Correlation properties of random fields \eqref{eta_xi-def-new} result from equality to zero of function $\zeta(x)$ mean value and formulas for its binary correlators, which we \textit{declare} in the following form,
\begin{subequations}\label{Basic_corrs(x)}
	\begin{align}
		\label{zeta_zeta-corr(x)}
		& \av{{\zeta}(x){\zeta}(x')} = \bm{\Xi}^2 W(|x-x'|)\theta(L/2-|x|)\theta(L/2-|x'|)\ ,\\[6pt]
		\label{zeta_zeta*-corr(x)}
		& \av{{\zeta}^*(x){\zeta}(x')} = |\bm{\Xi}|^2 W(|x-x'|)\theta(L/2-|x|)\theta(L/2-|x'|)\ .
	\end{align}
\end{subequations}
Hereinafter, angular brackets $\av{...}$ will stand for the averaging over realizations of random function $\zeta(x)$, complex parameter $\bm{\Xi}=\sigma_R+i\sigma_I$ denotes root-mean-square variance of this function. We assume that  correlation function $W(x)$  is specified on the entire axis $x$, and it is real, even, normalized to unity at the maximum at point $x=0$, and decaying rather rapidly to negligible values within the interval $|\Delta x|\sim r_c$ (the correlation radius).

The dependence of function $W$ in formulas \eqref{Basic_corrs(x)} on modulus $|x-x'|$ implies statistical homogeneity and statistical isotropy of the system under consideration. In the strict sense, such uniformity is lacking in systems of finite size. Yet, since it is easier to perform calculations with a one-coordinate correlation function than with a two-coordinate one, we have introduced the necessary $\theta$-factors into equations \eqref{Basic_corrs(x)} in order to account for finiteness of the support of function $\zeta(x)$ which we regard as statistically homogeneous and isotropic on the entire axis $x$. For now on we will restrict ourselves only to correlator set \eqref{Basic_corrs(x)} which, firstly, is justified for Gaussian random processes and, besides, for random processes of pretty arbitrary statistics yet under condition only that the scattering associated with these functions can be thought of as weak ~\cite{LifGredPast88}.

Average values of random fields  \eqref{eta_xi-def-new} are obviously equal to zero,  $\av{\eta(x)}= \av{\xi_\pm(x)}=0$. The calculation of binary correlators, although it is not fundamentally difficult, is technically an awkward procedure.
Calculation details are given in \ref{Bin_corr-details}, and here we give only its main results.

The binary correlator of random field \eqref{eta-def-new} describing the ``forward'' scattering (i.\,e., the scattering  with small momentum transfer) under WS conditions is equal to
\begin{subequations}\label{Corrs_eta,xi}
	\begin{equation}\label{<eta.eta>}
		\av{\eta(x)\eta(y)}\approx \frac{1}{L_f}F_l(x-y)\ .
	\end{equation}
	Here $L_f$  is the forward scattering length whose reciprocal is given by the sum of expressions \eqref{1/Lf-1} and \eqref{1/Lf-2}. Function $F_l(x)$ has the form \eqref{F_l(x)}. Its real ``width'' is of the order of the mesoscopic interval of local spatial averaging ~$l$. So, on the scale of the ``macroscopic'' lengths of our problem, to which we refer, in particular, the dynamic scattering lengths, the length of SPP dissipative damping, and the length $L$ of disordered segment of the interface, all of which significantly exceed the wavelength and the correlation length, this function can be considered as a prelimit $\delta$-function.
	
	The binary correlator of random field $\xi_\pm(x)$ from Eq.~\eqref{xi-def-new} and its conjugate counterpart is given by the expression similar to \eqref{<eta.eta>},  where the inverse \textit{backscattering} length $L_b$ stands for the coefficient before quasi-delta factor $F_l(x-y)$. By analogy with inverse length $L_f$, it includes a couple of terms given by Eq.~\eqref{1/Lb-1} and Eq.~\eqref{1/Lb-2},
	\begin{align}\label{<xi_pm(x)xi_pm(y)*>}
		\av{\xi_\pm(x)\xi_\mp(y)}= \frac{1}{L_b}F_l(x-y)\ .
	\end{align}
\end{subequations}

The correlator of non-conjugate fields $\xi_\pm(x)$ and $\xi_\pm(y)$, as well as the correlators of the field $\eta(x)$ and any of the fields $\xi_\pm(y)$ under WS conditions are parametrically small in comparison with correlators ~\eqref{Corrs_eta,xi}, and therefore they are further neglected. The correlator of random fields with tilde signs, see Eq.~\eqref{Tilde_xi->xi}, coincides with the one given by \eqref{<xi_pm(x)xi_pm(y)*>} due to their different renormalization by dissipative exponential factors,
\begin{align}\label{<xi_pm(x)xi_pm(y)*>-tilde}
	\av{\widetilde{\xi}_\pm(x)\widetilde{\xi}_\mp(y)}= \frac{1}{L_b}F_l(x-y)\ .
\end{align}
%
\section{The general solution for the leaking field}
\label{Full_Sol}
%
We will seek emitted field $h(\mathbf{r})$ for all $-\infty<x<\infty$, considering, however, the interface region $\mathbb{L}$ with a perturbed surface impedance as the only source of this field for the entire upper half-space. The expression
\begin{equation}\label{Full_solution_h(r)}
	h(\mathbf{r})=\int\limits_{-\infty}^{\infty}\frac{dq}{2\pi}
	\widetilde{\mathcal{R}}(q)\exp\left[iq x+i\big(k^2-q^2\big)^{1/2}z\right]
\end{equation}
by construction meets  wave equation \eqref{Helmholtz_eq} in the free (upper) half-space and radiation conditions at the infinity, so it is a good starting point for finding the radiation field.

It follows from representation \eqref{h-BC} that trial field $H_0(\mathbf{r})$ also satisfies equation~\eqref{Helmholtz_eq}. However, it does not suffices to meet boundary condition \eqref{Impedance_BC}. As far as in fact this BC is to be fulfilled by the sum of fields $H_0(\mathbf{r})$ and $h(\mathbf{r})$, we are free to choose trial function $H_0(\mathbf{r})$ in region $x\in\mathbb{L}$ in the artificial form \eqref{H_0^in}, which do not certainly satisfy condition \eqref{Impedance_BC}, yet compensate the emergent mismatch of the exact BC by means of condition \eqref{psi(x)-BC_in_L} for function $h(\mathbf{r})$, which in fact is directly related to the BC for the trial function.

By inserting function $h(\mathbf{r})$ in the form \eqref{Full_solution_h(r)} into boundary condition \eqref{psi(x)-BC_in_L} and performing Fourier transformation of the resulting equation with respect to coordinate $x$ (formally, along the entire axis), we obtain the following integral equation for kernel $\widetilde{\mathcal{R}}(q)$,
\begin{align}\label{R(q)-mainEq}
	\Big[\zeta_0+\sqrt{1-(q/k)^2}\,\Big]\widetilde{\mathcal{R}}(q)
	\,+\int\limits_{-\infty}^{\infty}\frac{dq'}{2\pi} &
	\widetilde{\mathcal{R}}(q')\widetilde{\zeta}(q-q')=
\notag\\
    = & -\int\limits_{-\infty}^{\infty}\frac{dq'}{2\pi}
	\widetilde{\zeta}(q-q')\widetilde{\mathcal{H}}^{(int)}_0(q')\ .
\end{align}
Tildes over symbols here denote the Fourier transform of the corresponding function with respect to variable $x$. Equations of this type were previously analyzed in Ref.~\cite{Depine92}, yet only the solutions obtained in the lowest orders of perturbation theory with respect to the integral term in the left-hand side of this equation were considered. Here we intend to obtain the solution valid regardless of which of the terms in the left-hand side of Eq.~\eqref{R(q)-mainEq} is the key one when constructing the perturbation theory.

We can write down the formal analytical solution to Eq.~\eqref{R(q)-mainEq} if the real part of the surface impedance, which accounts for dissipative processes in the metal, is nonzero ($\zeta_0'\neq 0$). In this case, the factor in square brackets in the left-hand side of this equation does not vanish on the real $q$-axis. By dividing both sides of Eq.~\eqref{R(q)-mainEq} by this factor we can rewrite it in the form of standard Fredholm integral equation of the second kind  \cite{bib:KolmogFomin68, bib:Courant_Hilbert66},
\begin{subequations}\label{Fredholm}
	\begin{align}\label{Fredholm_eq}
		\widetilde{\mathcal{R}}(q)+
		\int\limits_{-\infty}^{\infty}\frac{dq'}{2\pi}\mathcal{L}(q,q')\widetilde{\mathcal{R}}(q')
		=-\int\limits_{-\infty}^{\infty}\frac{dq'}{2\pi}
		\mathcal{L}(q,q')\widetilde{\mathcal{H}}^{(in)}_0(q')\ .
	\end{align}
The kernel of the integral operator in the lhs of  Eq.~\eqref{Fredholm_eq} is as follows,
	\begin{equation}\label{Kernel_Fredholm_eq}
		\mathcal{L}(q,q')=\left[\zeta_0+\sqrt{1-(q/k)^2}\right]^{-1}\widetilde{\zeta}(q-q')\ .
	\end{equation}
\end{subequations}
Under our assumptions about the impedance it is the Hilbert-Schmidt kernel, which guarantees the unique solvability of this equation.

The solution to Eq.~\eqref{Fredholm_eq} can be written in the following operator form,
\begin{equation}\label{R(B)R(SPP)-oper_form}
	\widetilde{\mathcal{R}}(q)=
	-\int\limits_{-\infty}^{\infty}\frac{dq'}{2\pi}\bra{q}\big(\hat{1}+
	\hat{\mathcal{L}}\big)^{-1}
	\hat{\mathcal{L}}\ket{q'}\widetilde{\mathcal{H}}^{(int)}_0(q')\ .
\end{equation}
We use Dirac notations for the operator matrix elements by means of \textit{bra}- and \textit{ket} -vectors. The operator $\hat{\mathcal{L}}$ with matrix elements \eqref{Kernel_Fredholm_eq} can be presented as a product of two operators, $\hat{G}^{(CP)}$ and $\hat{\zeta}_L$, which are given by the following matrix elements in momentum representation,
\begin{subequations}\label{Oper_notations}
	\begin{align}
		\label{hatG(S)}
		& \bra{q}\hat{G}^{(CP)}\ket{q'} =
		\left[\zeta_0+\sqrt{1-(q/k)^2}\right]^{-1}
		2\pi\delta(q-q')\ ,\\
		\label{hat_tilde_zeta}
		& \bra{q}\hat{\zeta}_L\ket{q'} =\widetilde{\zeta}(q-q')\ .
	\end{align}
\end{subequations}
The first of these operators, $\hat{G}^{(CP)}$, is a propagator of some wave formation that we will further refer to as the \emph{composite plasmon} (the meaning of this term is explained below, see formulas \eqref{Int_G(SPP)(x,x')}). The second operator factor,  $\hat{\zeta}_L$, is referred to as the impedance perturbation operator. The multiplication of operator matrices is defined in a~standard way, as the integral of the form
$\int_{-\infty}^{\infty}(dq/2\pi)\dots\ket{q}\bra{q}\dots$\,. The action of operators on state vectors $\ket{\cdot}$ with \textit{complex} argument  $\mathcal{K}$  is understood as follows,
\begin{align}\label{Oper->ket_vec}
	\bra{q}\hat{\mathcal{B}}\ket{\mathcal{K}}\stackrel {\text{def}}{=}
	\int\limits_{-\infty}^{\infty}\frac{dq'}{2\pi}\bra{q}\hat{\mathcal{B}}\ket{q'}
	\Delta_L(q'-\mathcal{K})\ .
\end{align}
Here, $\Delta_L$ denotes  the \textit{prelimit} $2\pi\delta$-function,
\begin{equation}\label{Underlimit_delta}
	\Delta_L(\kappa)=\int\limits_{-L/2}^{L/2}dx\,e^{\pm i\kappa x}
	=\frac{\sin(\kappa L/2)}{\kappa/2}\quad\left(\xrightarrow[L\to\infty]{}2\pi\delta(\kappa)
	\quad\text{if}\quad \mathrm{Im}\,\kappa=0\right)\ .
\end{equation}
This definition is due to the finiteness of the interval in which  function $\zeta(x)$ is defined. The representation  Eq.~\eqref{Oper->ket_vec} allows one to avoid formal mathematical difficulties in the transition from the space of real wave numbers, normally used in expansions into Fourier integrals, to complex wave numbers one of which, particularly in our problem, is the plasmon--polariton wave parameter~$k_{spp}$.

Substitution of  solution \eqref{R(B)R(SPP)-oper_form} into formula \eqref{Full_solution_h(r)} provides an opportunity to find with any prescribed accuracy the leaky field at arbitrary point of the half-space $z> 0$ and to analyze the transmitted, reflected and emitted into the upper half-plane fields into which the incident plasmon--polariton can be scattered by the defective interface segment. In fact, it is difficult to perform the explicit integration over $q$ in formula \eqref{Full_solution_h(r)} in view of the nontrivial functional structure of the scattering amplitude \eqref{R(B)R(SPP)-oper_form}. Yet, the problem can be significantly simplified in limiting cases allowing for the expansion of the inverse operator in the right-hand part of Eq.~\eqref{R(B)R(SPP)-oper_form} into some functional power series, whose members can be interpreted in terms of the multiplicity of scattering due to impedance perturbations.

To perform the required expansion correctly, it is necessary to define the operator~$\hat{\mathcal{L}}$ norm, whose detailed calculation is given in \ref{App_A}. Based on the results given there, we will perform statistical averaging of the scattered field in the limiting cases of weak and strong mixing of surface and bulk scattered modes. Yet, previously we give the interpretation of operator $\hat{G}^{(CP)}$ which is included in perturbation operator $\hat{\mathcal{L}}$.

Formal representation of the operator standing between  \textit{bra}- and \textit{ket}-vectors in Eq.~\eqref{R(B)R(SPP)-oper_form} as a sum of the operator power series  (at $\|\hat{\mathcal{L}}\|<1$ such a representation is mathematically rigorous) allows, at first glance, to interpret (in diagrammatic language) the operator  $\hat{\zeta}_L$ as a vertex operator, while matrix elements \eqref{hatG(S)} as the unperturbed propagators of scattered wave excitations (presumably, plasmon polaritons~\cite{bib:Tejeira05,Many-many08}). This interpretation of operator $\hat{G}^{(CP)}$ is not, however, quite rigorous, as indicated by the authors of the above cited works, and may be considered satisfactory only in the asymptotic sense.

\begin{figure}[h!!]
	\centering
	\scalebox{.8}[.8]{\includegraphics{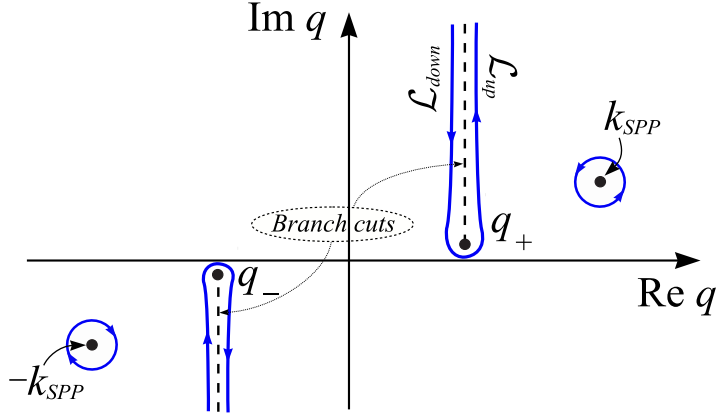}}
	\caption{Deformation of the integration contour in formula~\eqref{G(SPP)(x,x')}.
		\hfill\label{fig2-old}}
\end{figure}

To shed light on the true physical meaning of this operator, we transform its matrix elements from momentum to coordinate representation. To do this, it is necessary to calculate the integral
\begin{equation}\label{G(SPP)(x,x')}
	\bra{x}\hat{G}^{(CP)}\ket{x'} =\int\limits_{-\infty}^{\infty}\frac{dq}{2\pi}
	\frac{e^{iq(x-x')}}{\zeta_0+\sqrt{1-(q/k)^2}}\ ,
\end{equation}
whose integrand has singularities of two types in the plane of complex $q$ variable, specifically, the poles at points $\pm k_{spp}$ and branch points $q_{\pm}=\pm (k+i0)$ (see~Fig.~\ref{fig2-old}).
In the figure, the displacement of the branch points along the vertical axis, which sets the rule for bypassing them during contour integration, is in practice ensured by some attenuation (up to infinitesimal), which is always present in real systems.

Since integral \eqref{G(SPP)(x,x')} depends actually on the \emph{modulus} of difference $x-x'$,  it can be calculated by closing the integration contour, for example, in the upper half-plane of complex $q$. In doing so, we in fact specify the choice of the root sign in the definition of the plasmon--polariton wave number in formula \eqref{RightLeft_SPP}. Specifically, by setting  $\zeta''_0=\mathrm{Im}\,\zeta_0<0$ and $\zeta'_0=\mathrm{Re}\,\zeta_0>0$, we will assume that $\mathrm{Im}\,\sqrt{1-\zeta_0^2}>0$.  As a result, integral~\eqref{G(SPP)(x,x')} may be represented as a sum of two terms which result from the contributions of the pole at $q=k_{spp}$ and the cut edges drawn in the upper half-plane, viz.,
\begin{subequations}\label{Int_G(SPP)(x,x')}
\begin{align}
\label{Int_G(SPP)_pole+branch}
  &	\bra{x}\hat{G}^{(CP)}\ket{x'} =
   {G}^{(CP)}_{pole}(x,x')+
   {G}^{(CP)}_{br-c}(x,x')\ ,\\
\label{Int_G(SPP)_pole}
  &	{G}^{(CP)}_{pole}(x,x')=
   \frac{ik^2\zeta_0}{k_{spp}}
   \exp\Big(ik_{spp}|x-x'|\Big)=
\notag\\
  & \phantom{ \exp\Big(ik_{spp}|x-x'|\Big)}
  =\frac{ik\zeta_0}{\sqrt{1-\zeta_0^2}}
   \exp\Big(ik\sqrt{1-\zeta_0^2}|x-x'|\Big)\ ,\\
\label{Int_G(SPP)_branch}
  & {G}^{(CP)}_{br-c}(x,x')= \frac{-ik}{\pi}
   e^{ik|x-x'|}\,\mathcal{G}(x-x')\ ,
\end{align}
\end{subequations}
where
$$
   \mathcal{G}(x-x') = \int\limits_0^{\infty}du\,e^{-u k|x-x'|}
   \frac{\sqrt{u(u-2i)}}{u(u-2i)-\zeta_0^2}\ .
$$

The first, the ``pole" term in Eq.~\eqref{Int_G(SPP)_pole+branch} is the Green's function of the plasmon--polariton, the truly \textit{surface} EM wave which decays in the direction of propagation provided the underlying surface possesses some dissipative properties. The second, the cut-related term also is of wave nature, yet the amplitude of this wave at large distances from the source (from point $x'$) decreases proportionally to $|x-x'|^{-3/2}$, regardless of dissipation level in the metal. This wave, thus, does not belong to a class of surface waves. Spatial harmonics of the field comprised in term \eqref{Int_G(SPP)_branch}, as can be seen from its structure, are not localized at the conductor surface and can be considered as forming the ``radiation part'' of the scattered field. The most suitable candidate for the role of a leaky field are the so-called Norton waves, described in detail in papers~\cite{Norton36,Norton37,Norton37_2}. The presence in propagator \eqref{Int_G(SPP)_pole+branch} of both the plasmon--polariton and the ``leaking'' component of the scattered field can be the reason to characterize the wave states described by this propagator as the \emph{composite plasmons} (CPs).
%
\section{On the scattering amplitude asymptotics}
\label{Asymptota}
%
Scattering amplitude \eqref{R(B)R(SPP)-oper_form} is represented as a linear combination of the matrix elements of complex-valued operator
\begin{equation}\label{T=(1+L)^[-1]L}
	{\hat{\mathcal{T}}=\big(\hat{1}+\hat{\mathcal{L}}\big)^{-1}\hat{\mathcal{L}}}\ ,
\end{equation}
which plays the role of a propagator of initial states described by \textit{ket}-vectors $\ket{q'}$ into the final state of scattering, which corresponds to the \textit{bra}-vector  $\bra{q}$. In this notation, the operator $\hat{\mathcal{L}}$ defined by matrix elements \eqref{Kernel_Fredholm_eq} can be interpreted as the operator of a \emph{single} scattering of the modes with arbitrary $x$-components of the momentum from the ``initial'' state corresponding to the right mode index to the ``final'' (left-standing) mode which is prescribed by propagator~\eqref{hatG(S)}. The scattering comes about through the intermediate states with propagators \eqref{Int_G(SPP)(x,x')}, and therefore we will refer to it as the CP-mediated scattering.

The intensity of the scattering produced by the operator~$\hat{\mathcal{L}}$ is naturally characterized by the properly determined \textit{norm} of this operator, which should be defined from physical considerations taking account of the full set of allowable initial and final scattering states. Depending on the norm value, we will consider two limiting cases: weak and strong mixing of surface (SPP-like) and bulk (Norton-like)  components of the field $H_0^{(int)}(\mathbf{r})$ considered only as a trial but not a true magnetic field within non-uniform interval  $\mathbb{L}$.

In our model, the trial field on the entire $x$-axis is chosen to have fixed SPP-like dependence on the coordinate~$z$, which coincides with the dependence on this coordinate of the unperturbed SPP at the left and right edges of the disordered region. This choice of the trial field gives us the reasoning to regard operator $\hat{\mathcal{L}}$, from technical side, as the operator resulting in the plasma wave scattering strictly in one instead of in two dimensions.  This operator is non-Hermitian because it contains both the surface (SPP-like) and the bulk field harmonics through which the SPP energy is tranferred into the upper half-space. The non-Hermitian nature of $\hat{\mathcal{L}}$ operator is also provided by energy loss in the metal (the real part of the impedance) and the plasmon--polariton propagation/reflection through/from segment $\mathbb{L}$ which is at both ends open. In what follows, we will refer to operator $\hat{\mathcal{L}}$ as the \textit{mixing} operator for one-dimensional harmonics of the trial field (which has a purely surface nature) and the outward-radiated harmonics with nonzero $z$-components of the wave vector.
%
\subsection{Single CP-mediated scattering}
%
The multiplicative structure of operator $\hat{\mathcal{L}}=\hat{G}^{(CP)}\hat{\zeta}_L$ allows one to interpret its action on the field $\mathcal{H}^{(int)}_0(x)$ in the~following way. Vertex operator $\hat{\zeta}_L$  transforms each of the harmonics of an arbitrary field subject to scattering into the sum of freely-propagating SPP wave, whose propagator is given by Eq.~\eqref{Int_G(SPP)_pole}, and a bunch of additional waves with non-dissipative power-law attenuation along  $z$-coordinate from an arbitrary point on the axis $x$.  If each of these $x$-points is considered as a \textit{local} source for the leaking field, then the structure of the latter is obvious: it consists of plasmon--polariton component propagating strictly along the surface and the leaking component propagating into the upper half-space in the form of Norton waves.

The structure of operator \eqref{T=(1+L)^[-1]L} indicates that generally it describes the \textit{multiple} scattering of an arbitrary field incident on the disordered surface region through the intermediate CP states.  In this case, the scattered field intensity distribution between purely surface (SPP) component and its leaking (quasi-Nortonian) component in each separate scattering act can be easily determined from formulas ~\eqref{Int_G(SPP)(x,x')}.

We will refer to CP-mediated scattering of a seed field in formula \eqref{R(B)R(SPP)-oper_form} as a weak one or, equivalently, as a \textit{single} scattering if the condition is met
\begin{equation}\label{Weak_intermixing}
	\Av{\|\hat{\mathcal{L}}\|^2}\ll 1\ .
\end{equation}
By expanding inverse operator on the right-hand part of formula \eqref{R(B)R(SPP)-oper_form} into an operator power series and keeping only two first terms of this series we can approximate the scattering amplitude  $\widetilde{\mathcal{R}}(q)$ by the expression as follows,
\begin{align}\label{R(B)R(S)-|L|<<1}
	\widetilde{\mathcal{R}}(q)\approx -\int\limits_{-\infty}^{\infty}\frac{dq'}{2\pi}
	\bra{q}\big(\hat{1}-\hat{\mathcal{L}}\big)
	\hat{\mathcal{L}}\ket{q'}\widetilde{\mathcal{H}}^{(int)}_0(q')\ .
\end{align}
Here, in addition to the linear-in-$\hat{\mathcal{L}}$ term we keep the term quadratic in $\hat{\mathcal{L}}$ in order to be able, if necessary, to determine the nonzero contribution of scattering to the \emph{mass operator} (spectrum correction), which in the statistical theory of wave scattering usually determines the \emph{mean} field. In the latter, the terms linear in $\hat{\mathcal{L}}$ normally vanish.
%
\subsection{Multiple CP-mediated scattering}
%
The problem cannot be limited to the single scattering and, accordingly, the mixing of surface and bulk scattered modes cannot be considered weak if the norm of the operator $\hat{\mathcal{L}}$ approaches unity in order of magnitude or, especially, if it exceeds unity value. As the limit of strong coupling of surface and bulk scattered modes (or, equivalently, strong CP-mediated scattering), it is natural to consider the case where the inequality holds
\begin{equation}\label{Strong_intermixing}
	\Av{\|\hat{\mathcal{L}}\|^2}\gg 1\ .
\end{equation}
From estimates \eqref{Square_norm_L1><} and \eqref{Norm_estim-gen} it follows that such an inequality is realistic even for small fluctuations of the impedance provided the dissipation in metal is small enough, $\zeta'_0\ll 1$. But it may well be fulfilled even in metals with high level of dissipation. This, however, requires the impedance dispersion to be sufficiently large ($|\bm{\Xi}|^2\gg 1$).

When inequality \eqref{Strong_intermixing} holds true, it is reasonable to re-write, by simple algebraic manipulations with operator \eqref{T=(1+L)^[-1]L}, the formula for the scattering amplitude in the following form,
\begin{align}
	\label{R(B)R(SPP)-inverse_L}
	\widetilde{\mathcal{R}}(q) & =-\int\limits_{-\infty}^{\infty}\frac{dq'}{2\pi}
	\bra{q}\big(\hat{1}+\hat{\mathcal{L}}\big)^{-1}
	\hat{\mathcal{L}}\ket{q'}\,\widetilde{\mathcal{H}}^{(in)}_0(q')
	\notag\\
	& = -\widetilde{\mathcal{H}}^{(in)}_0(q)
	+\int\limits_{-\infty}^{\infty}\frac{dq'}{2\pi}
	\bra{q}\big(\hat{1}+\hat{\mathcal{L}}\big)^{-1}\ket{q'}\,
	\widetilde{\mathcal{H}}^{(int)}_0(q')\ .
\end{align}
Both lines in Eq.~\eqref{R(B)R(SPP)-inverse_L} are equivalent. Yet, the expression in the first line is more convenient to use when considering the limiting case \eqref{Weak_intermixing} of weak CP-mediated scattering whereas the second expression is better for cases where $\|\hat{\mathcal{L}}\|^2\gg 1$ corresponding to strong intermixing of SPP and leaking components of the scattered field. The latter occasion deserves especial consideration, so we will return to it thoroughly in a separate publication. In the present article, we will focus on the limiting case of weak CP-mediated scattering, when the scattered bulk and surface harmonics are intermixed weakly.
%
\section{Scattering pattern of a surface polariton under weak \\ CP-mediated scattering}
\label{ScattDiagr}
%
As shown in \ref{App_A} (see also Eqs.~\eqref{Int_G(SPP)(x,x')}), operator $\hat{\mathcal{L}}$  can be interpreted as a mixer of two fundamentally different scattering channels: the one-dimensional SPP channel where the end result of the SPP scattering are also surface plasmons, and the leakage channel with the outcome in the form of quasi-isotropic leaking (emitted) waves. Which of these channels dominates in the scattering process depends on the ratio of the corresponding operator norms, see formulas  \eqref{Square_norm_L1><} and \eqref{Norm_estim-gen}. At the same time, regardless of the scattering rates in both of the above channels, we can discuss both weak and strong scattering in terms of the \textit{total} norm of the operator $\hat{\mathcal{L}}$. Further in this article, we will consider in more detail the case $\|\hat{\mathcal{L}}\|\ll 1$, since just in this case the  sharp and hitherto unexplained anisotropy of the emitted radiation is manifested most evidently.

At first, let us find out which initial parameters of our problem correspond to the fulfillment of inequality ${\|\hat{\mathcal{L}}\|\ll 1}$, that naturally supposes simultaneous fulfillment of two inequalities, namely,
\begin{equation}\label{|L|_spp<<1,|L|_rad<<1}
	\Av{\|\hat{\mathcal{L}}\|^2}^{(pole)}\ll 1\ ,\qquad
	\Av{\|\hat{\mathcal{L}}\|^2}^{(rad)}\ll 1\ ,
\end{equation}
each of which is an estimate of the scattering rate of the trial field into a particular channel, either SPP or the radiation one, see \ref{App_A}. Assuming, for definiteness, $|\zeta_0''|\sim 1$  the first of these inequalities can be written as
\begin{subequations}\label{|L|<<1-2_ineqs}
	\begin{equation}\label{|L|<<1-1st}
		|\Xi|^2\frac{1}{\zeta_0'}k\widetilde{W}(k'_{spp})\ll 1\ ,
	\end{equation}
while the second, the ``radiation'' inequality, reduces to
	\begin{equation}\label{|L|<<1-2nd}
		|\Xi|^2\min(kr_c, 1)\ll 1\ .
	\end{equation}
\end{subequations}
If $kr_c\ll 1$, then $\widetilde{W}(k'_{spp})\approx r_c$  and taking into account that in good metals $\zeta_0'\ll 1$, we obtain that condition ${\|\hat{\mathcal{L}}\|\ll 1}$ is equivalent to inequality
\begin{subequations}\label{|L|<<1-krc><1}
	\begin{equation}\label{|L|<<1=>Weak_krc<<1}
		|\Xi|^2\frac{kr_c}{\zeta_0'}\ll 1 \qquad\quad(kr_c\ll 1)\ .
	\end{equation}
In this case, the dominant scattering channel is the SPP one.

If $kr_c\gg 1$, then correlation function $\widetilde{W}(k'_{spp})$ takes small values in comparison with $r_c$, and starting from some value of wave number~$k$, such that $k\widetilde{W}(k'_{spp})$ becomes by the order of magnitude comparable with~$\zeta_0'$, the radiation channel appears to dominate and the weakness condition for the CP-mediated scattering is reduced to inequality
	\begin{equation}\label{|L|<<1=>Weak_krc>>1}
		|\Xi|^2\ll 1\ .
	\end{equation}
\end{subequations}
Interesting is to note that for small-scale impedance fluctuations ($kr_c\ll 1$) the scattering can remain weak even at rather large fluctuation amplitude, when $|\Xi|^2\gtrsim 1$. This, however, fully agree with quantum-mechanical interpretation of the scattering amplitude as the \textit{integral} rather than local scattering characteristic.

Let us now compare the conditions \eqref{|L|<<1-krc><1} of weak CP-mediated scattering with the condition of weak \textit{dynamic} scattering formulated previously in the form of the requirement for the smoothness of quasi-amplitude factors $\pi(x)$ and $\gamma(x)$ in representation~\eqref{H^in->pi_gamma}. The latter condition, in fact,  reduces to the requirement that spatial decrement $L^{-1}_f$ must be small in comparison with wave parameter~$k$ in the free half-space. From \eqref{1/Lf-1} this decrement can be estimated as
\begin{equation}\label{1/Lf-estim}
	\frac{1}{L_f}\sim k|\Xi|^2kr_c\zeta_0\ .
\end{equation}
By comparing this estimate with the expression standing in the left-hand side of inequality \eqref{|L|<<1=>Weak_krc<<1} we come to a conclusion that for $kr_c\ll 1$ the condition for weak CP-mediated scattering is not always satisfied. The dynamic (the one particle-type, in terms of quantum mechanics) scattering, which determines the spectrum of SPP harmonics,  is conventionally regarded as weak if the inequality holds $L_f^{-1}\ll k$. At the same time, in order for the CP-mediated scattering be considered weak the inequality should be met ${L_f^{-1}\ll \zeta'_0 k}$. For $kr_c\gg 1$, when correlation function $\widetilde{W}(k'_{spp})$ is parametrically small quantity, the criteria for the dynamic and the CP-mediated scattering to be weak simultaneously are not mutually related. This means that these two types of scattering can be taken into account independently of each other.

Assuming that inequalities \eqref{|L|<<1-2_ineqs} are simultaneously satisfied and substituting scattering amplitude  \eqref{R(B)R(S)-|L|<<1} with only linear in $\hat{\mathcal{L}}$ term left into Eq.~\eqref{Full_solution_h(r)}, we obtain the following expression for the radiation field in the domain~$z>0$:
\begin{align}\label{Scatt_field-weak}
	h(\mathbf{r}) &\approx -ik\int\limits_{-\infty}^{\infty}\frac{dq}{2\pi}\exp\left(iqx+i\sqrt{k^2-q^2}z\right)
	\int\limits_{-\infty}^{\infty}\frac{dq'}{2\pi}
	\hat{\mathcal{L}}(q,q')\widetilde{\mathcal{H}}^{(in)}_0(q')
	\notag\\
	& = -ik \int\limits_{-\infty}^{\infty}\frac{dq}{2\pi}
	\frac{\exp\left(iqx+i\sqrt{k^2-q^2}z\right)}{\zeta_0+\sqrt{1-(q/k)^2}}
	\int\limits_{-\infty}^{\infty}\frac{dq'}{2\pi}
	\widetilde{\zeta}(q-q') \widetilde{\mathcal{H}}^{(in)}_0(q')\ .
\end{align}
Using Eq.~\eqref{Scatt_field-weak}, we can analyze the scattering diagram in terms of the average (over realizations of random function $\zeta(x)$) intensity of the field  $h(\mathbf{r})$. For the averaging procedure to be physically reasonable we will also assume the fulfillment of inequality
\begin{equation}\label{r_c<<L}
	r_c\ll L\ .
\end{equation}
In addition, we will consider the intensity of the radiated field at large distances from the scattering object, viz., the disordered boundary segment $\mathbb{L}$,
\begin{equation}\label{Large_distance}
	|\mathbf{r}|=\sqrt{x^2+z^2}\gg L\ .
\end{equation}
%
\subsection{The average intensity of the leaking field}
\label{Av_intensity}
%
The average (over realizations of $\zeta(x)$ and over time) intensity of the field \eqref{Scatt_field-weak} is represented as
\begin{subequations}\label{Intens_RadField}
\begin{align}
\label{Intens_gen}
	\Av{|h(\mathbf{r})|^2}=
	\frac{1}{2}\iint\limits_{-\infty}^{\quad\infty}\frac{dqdq'}{(2\pi)^2}
	\exp\bigg\{i(q-q')x+i\Big[ & \sqrt{k^2-q^2}-\left(\sqrt{k^2-q'^2}\right)^*\Big]z\bigg\}\times
\notag\\
   & \quad\quad \times\AV{\widetilde{\mathcal{R}}(q)\widetilde{\mathcal{R}}^*(q')},
\end{align}
where
\begin{align}
\label{R(q)R*(q')}
	\AV{\widetilde{\mathcal{R}}(q)\widetilde{\mathcal{R}}^*(q')}=
	& \iint\limits_{-\infty}^{\quad\infty}\frac{dq_1dq_2}{(2\pi)^2}
	\bigg<\bra{q}\big(1+\hat{\mathcal{L}}\big)^{-1}\hat{\mathcal{L}}\ket{q_1}
    \left[\bra{q'}\big(1+\hat{\mathcal{L}}\big)^{-1}\hat{\mathcal{L}}\ket{q_2}\right]^* \times
\notag\\
   & \phantom{\quad \Big<\bra{q}\big(1+\hat{\mathcal{L}}\big)^{-1}\hat{\mathcal{L}}\ket{q_1}}
    \times\widetilde{\mathcal{H}}^{(in)}_0(q_1)\left[\widetilde{\mathcal{H}}^{(in)}_0(q_2)\right]^*\bigg>\ .
\end{align}
\end{subequations}
The expression for correlator \eqref{R(q)R*(q')} can be substantially simplified if we take into account that random functions in operator $\hat{\mathcal{L}}$ matrix elements and in trial functions  $\widetilde{\mathcal{H}}^{(in)}_0(q_{1,2})$ are fundamentally different. Function $\widetilde{\mathcal{H}}^{(in)}_0(q)$ is actually a functional of  smoothed random fields \eqref{eta_xi-def-new} while the operator $\hat{\mathcal{L}}$ kernel (see Eq.~\eqref{Kernel_Fredholm_eq}) depends on the random part of the impedance in its bare form. The result of mutual  ``pairing'' of random functions contained in $\hat{\mathcal{L}}$ and in $\widetilde{\mathcal{H}}^{(in)}_0(q)$ appear to be subjected to extra averaging over space interval $l$ satisfying conditions \eqref{lambda<l<L_sc-new},  at which, due to the uncompensation of ``rapid'' phase factors, the result becomes parametrically reduced. If we neglect such depressed pairings, matrix elements of propagator  \eqref{T=(1+L)^[-1]L}, on the one hand, and the trial functions in  Eq.~\eqref{R(q)R*(q')}, on the other, can be averaged separately, thus resulting in representation
\begin{align}
\label{|h(r)|^2-appr}
	\AV{\widetilde{\mathcal{R}}(q)\widetilde{\mathcal{R}}^*(q')} \approx
	\iint\limits_{-\infty}^{\quad\infty}\frac{dq_1dq_2}{(2\pi)^2}
	\bm{\mathcal{K}}_1(q,q'|q_1,q_2)
	\bm{\mathcal{K}}_2(q_1,q_2) \ .
\end{align}
Under conditions of weak coupling between SPP and the radiation scattering channels, i.\,e., when inequality holds ${\Av{\|\hat{\mathcal{L}}\|}^2\ll 1}$, the inverse operators $\big(1+\hat{\mathcal{L}}\big)^{-1}$ in the first of the correlators in Eq.~\eqref{|h(r)|^2-appr}, namely,
\begin{equation}\label{1st_corr_orig}
  \bm{\mathcal{K}}_1(q,q'|q_1,q_2)=\Big< \bra{q}\big(1+\hat{\mathcal{L}}\big)^{-1}\hat{\mathcal{L}}\ket{q_1}
	\left[\bra{q'}\big(1+\hat{\mathcal{L}}\big)^{-1}\hat{\mathcal{L}}\ket{q_2}\right]^*\Big>\ ,
\end{equation}
can be replaced with unity. Then this correlator can be easily calculated to
\begin{subequations}\label{K_1,K_2-defs}
	\begin{align}\label{1st_corr}
		\bm{\mathcal{K}}_1(q,q'|q_1,q_2)=\, & \frac{|\bm{\Xi}|^2}{\left[\zeta_0+\sqrt{1-(q/k)^2}\right] \left[\zeta_0+\sqrt{1-(q'/k)^2}\right]^*} \times
\notag\\
	&\qquad\times\int\limits_{-\infty}^{\infty}\frac{ds}{2\pi}\widetilde{W}(s)\Delta_L(s-q+q_1)\Delta_L(s-q'+q_2)
	\approx
\notag\\
    \approx\, & |\bm{\Xi}|^2\frac{\widetilde{W}(q-q_1)\Delta_L(q-q_1-q'+q_2)}{\left[\zeta_0+\sqrt{1-(q/k)^2}\right] \left[\zeta_0+\sqrt{1-(q'/k)^2}\right]^*}\ .
\end{align}

The second correlator in Eq.~\eqref{|h(r)|^2-appr}, $\bm{\mathcal{K}}_2(q_1,q_2)=\AV{\widetilde{\mathcal{H}}^{(in)}_0(q_1)\left[\widetilde{\mathcal{H}}^{(in)}_0(q_2)\right]^*}$, under WS conditions \eqref{lambda<l<L_sc-new} can be written as
\begin{align}\label{2nd_corr}
	\bm{\mathcal{K}}_2(q_1,q_2)
	\approx \iint_{\mathbb{L}}dx_1dx_2
	& \Big[\AV{\pi(x_1)\pi^*(x_2)}e^{-i(q_1-\varkappa'_{s})x_1+i(q_2-\varkappa'_{s})x_2} +
\notag\\
	& +\AV{\gamma(x_1)\gamma^*(x_2)}e^{-i(q_1+\varkappa'_{s})x_1+i(q_2+\varkappa'_{s})x_2}+
\notag\\
    & + i\AV{\pi(x_1)\gamma^*(x_2)}e^{-i(q_1-\varkappa'_{s})x_1+i(q_2+\varkappa'_{s})x_2} -
\notag\\
	& -i\AV{\gamma(x_1)\pi^*(x_2)}e^{-i(q_1+\varkappa'_{s})x_1+i(q_2-\varkappa'_{s})x_2} \Big]\ .
	\end{align}
\end{subequations}
By integrating the product of correlators  \eqref{1st_corr} and \eqref{2nd_corr} over $q_1$ and $q_2$ and then returning to the coordinate representation for correlation function $W(x)$ we get the expression as follows for the average radiated intensity,
\begin{align}\label{Int_q1q2_h(r)}
	\Av{|h(\mathbf{r})|^2} \approx\,  &
	\frac{|\bm{\Xi}|^2}{2}\iint\limits_{-\infty}^{\quad\infty}\frac{dqdq'}{(2\pi)^2}
	\tfrac{\exp\Big\{i(q-q')x+i\Big[\sqrt{k^2-q^2}-\left(\sqrt{k^2-q'^2}\right)^*\Big]z\Big\}}
	{\left[\zeta_0+\sqrt{1-(q/k)^2}\right] \left[\zeta_0+\sqrt{1-(q'/k)^2}\right]^*} J(q,q') ,
\end{align}
where
\begin{align}\label{J(q,q')}
J(q,q')
& = \iint_{\mathbb{L}}dx_1dx_2 W(x_1-x_2)e^{-iqx_1+iq'x_2} \times
\notag\\
& \times \bigg[ \Av{\pi(x_1)\pi^*(x_2)}e^{i\varkappa'_{s}(x_1-x_2)}+
\Av{\gamma(x_1)\gamma^*(x_2)}e^{-i\varkappa'_{s}(x_1-x_2)} +
\notag\\
& +i\Av{\pi(x_1)\gamma^*(x_2)}e^{i\varkappa'_{s}(x_1+x_2)}
-i\Av{\gamma(x_1)\pi^*(x_2)}e^{-i\varkappa'_{s}(x_1+x_2)}\bigg] .
\end{align}

To make expression \eqref{J(q,q')} more simple for further manipulations, it is natural to change the variables in integral \eqref{J(q,q')} to the new ones, namely, $\delta=(x_1-x_2)$ and $X = (x_1+x_2)/2$. The domain of integration becomes now more intricate, specifically, it is rhombus-shaped instead of the initial square. But since all two-point correlators in square brackets, in view of conditions \eqref{lambda<l<L_sc-new} and \eqref{r_c<<L}, are smooth functions in comparison with $W(\delta)$, we can extend the integral over $\delta$ to the whole interval $-L \leq \delta \leq L$. As a result, formula \eqref{J(q,q')} acquires the form
\begin{align}\label{J_simple}
J(q,q')
\approx & \int\limits_{-L}^{L}d\delta W(\delta )e^{i(q+q')\delta/2 }\int\limits_{-L/2}^{L/2}dX e^{-i\left( q-q'\right) X} \bigg[ \left\langle |\pi (X)|^{2}\right\rangle e^{i\varkappa'_{s}\delta }+
\notag \\
& + \left\langle |\gamma (X)|^{2}\right\rangle e^{-i\varkappa'_{s}\delta } +
i\left\langle \pi (X)\gamma^{\ast }(X)\right\rangle e^{2i\varkappa'_{s} X}
-i\left\langle \gamma (X)\pi ^{\ast }(X)\right\rangle e^{-2i\varkappa'_{s} X}\bigg] ,
\end{align}
whose advantage in comparison with Eq.~\eqref{J(q,q')} is that it requires to average \textit{single-coordinate} functions instead of two-coordinate ones. This is important from the mathematical point of view since the competition of an additional dimensional parameter $|x_1-x_2|$ with other length parameters, specifically, the length of the disordered region $\mathbb{L}$ as well as the scattering and  dissipation lengths, significantly complicates already tedious calculations.
%
\subsection{The technique of statistical averaging}
\label{Scatt_technique}
%
Since equations \eqref{pi_gamma-eqs-new2} are more simple in comparison with equations \eqref{pi_gamma-eqs-new} for non-tilde functions, we will perform the statistical averaging for functions \eqref{Tilde_pi_gamma} with tilde signs. For weak scattering, random functions $\eta(x)$ and $\widetilde{\xi}_{\pm}(x)$ can be considered as nearly Gaussian random processes \cite{LifGredPast88} with zero mean values and binary correlators \eqref{Corrs_eta,xi}, \eqref{<xi_pm(x)xi_pm(y)*>-tilde}. To average functions $|\pi(x)|^2$, $|\gamma(x)|^2$, $\pi(x)\gamma^*(x)$ and $\gamma(x)\pi^*(x)$, we use differential equations following directly from  Eqs.~\eqref{pi_gamma-eqs-new2},
\begin{subequations}\label{Auxil_eqs__pg}
	\begin{align}
		\label{|p|^2--eq}
		& \frac{d\,|\widetilde{\pi}(x)|^2}{dx}=-\xi_-(x)\widetilde{\gamma}(x)\widetilde{\pi}^*(x)-
		\xi_+(x)\widetilde{\pi}(x)\widetilde{\gamma}^*(x)\ ,\\
		\label{|g|^2--eq}
		& \frac{d\,|\widetilde{\gamma}(x)|^2}{dx}=-\xi_-(x)\widetilde{\gamma}(x)\widetilde{\pi}^*(x)-
		\xi_+(x)\widetilde{\pi}(x)\widetilde{\gamma}^*(x)\ ,\\
		\label{pg*--eq}
		& \frac{d}{dx}\big[\widetilde{\pi}(x)\widetilde{\gamma}^*(x)\big]=
		-2i\eta(x)\big[\widetilde{\pi}(x)\widetilde{\gamma}^*(x)\big]-
		\xi_-(x)\left[|\widetilde{\pi}(x)|^2+|\widetilde{\gamma}(x)|^2\right]\ ,\\
		\label{gp*--eq}
		& \frac{d}{dx}\big[\widetilde{\gamma}(x)\widetilde{\pi}^*(x)\big]=
		2i\eta(x)\big[\widetilde{\gamma}(x)\widetilde{\pi}^*(x)\big]-
		\xi_+(x)\left[|\widetilde{\pi}(x)|^2+|\widetilde{\gamma}(x)|^2\right]\ .
	\end{align}
\end{subequations}
For averaging the right-hand parts of Eqs.~\eqref{Auxil_eqs__pg} we apply Furutsu--Novikov formalism developed for functionals of Gaussian random processes \cite{Furutsu63,Novikov64}. Let us briefly recall here the basics of this formalism.

If $\mathbf{f}(x)$ is a zero-mean Gaussian random \textit{vector} field with arbitrary number of components $f_i(x)$ satisfying correlation equalities
\begin{equation}\label{<f_i.f_k>}
	\Av{f_i(x)f_k(x')}=\mathcal{D}_{ik}\delta(x-x')\ ,
\end{equation}
then for some given functional $\mathcal{Q}[\mathbf{f}]$ the equality holds true
\begin{align}
	\label{Furutsu-Novikov(+-)}
	\Av{f_i(x)\mathcal{Q}[\mathbf{f}]}=
	\sum_{k}\int_{\mathbb{L}}dx'\Av{f_i(x)f_k(x')}
	\AV{\frac{\delta \mathcal{Q}[\mathbf{f}]}{\delta f_k(x')}}=
	\sum_{k}\mathcal{D}_{ik}
	\AV{\frac{\delta \mathcal{Q}[\mathbf{f}]}{\delta f_k(x)}}\ .
\end{align}
Variational derivatives in the right-hand part of Eq.~\eqref{Furutsu-Novikov(+-)} can be calculated from the differential equation for functional $\mathcal{Q}[\mathbf{f}](x)$.

For example, let us average Eq.~\eqref{|p|^2--eq}. The averaging yields
\begin{equation}\label{<|pi|^2>-eq}
	\frac{d\,\Av{|\widetilde{\pi}(x)|^2}}{dx}=-\Av{\xi_-(x)\widetilde{\gamma}(x)\widetilde{\pi}^*(x)}-
	\Av{\xi_+(x)\widetilde{\pi}(x)\widetilde{\gamma}^*(x)}\ .
\end{equation}
Function $\widetilde{\gamma}(x)\widetilde{\pi}^*(x)$ in the rhs of Eq.~\eqref{<|pi|^2>-eq} obeys equation \eqref{gp*--eq}. By averaging the latter we obtain
\begin{equation}\label{<>}
	\frac{d}{dx}\Av{\big[\widetilde{\gamma}(x)\widetilde{\pi}^*(x)\big]}=
	2i\Av{\eta(x)\big[\widetilde{\gamma}(x)\widetilde{\pi}^*(x)\big]}-
	\Av{\xi_+(x)\left[|\widetilde{\pi}(x)|^2+|\widetilde{\gamma}(x)|^2\right]}\ .
\end{equation}
To calculate the averages in the rhs of Eq.~\eqref{<>} we may apply once again the rule~\eqref{Furutsu-Novikov(+-)}. By repeating this procedure with all equations \eqref{Auxil_eqs__pg} as many times as is necessary we arrive at the following set of equations for arisen mean quantities,
\begin{subequations}\label{<Auxil_eqs__pg>}
	\begin{align}
		\label{<|p|^2>--eq}
		& \frac{d\,\Av{|\widetilde{\pi}(x)|^2}}{dx}=-\frac{1}{L_b}\left[\Av{|\widetilde{\pi}(x)|^2}+ \Av{|\widetilde{\gamma}(x)|^2}\right]\ ,\\
		\label{<|g|^2>--eq}
		& \frac{d\,\Av{|\widetilde{\gamma}(x)|^2}}{dx}=-\frac{1}{L_b}\left[\Av{|\widetilde{\pi}(x)|^2}+ \Av{|\widetilde{\gamma}(x)|^2}\right]\ ,\\
		\label{<pg*>--eq}
		& \frac{d}{dx}\Av{\widetilde{\pi}(x)\widetilde{\gamma}^*(x)}=
		\left(\frac{2}{L_f}-\frac{1}{L_b}\right)\Av{\widetilde{\pi}(x)\widetilde{\gamma}^*(x)}\ ,\\
		\label{<gp*>--eq}
		& \frac{d}{dx}\Av{\widetilde{\gamma}(x)\widetilde{\pi}^*(x)}=
		\left(\frac{2}{L_f}-\frac{1}{L_b}\right)\Av{\widetilde{\gamma}(x)\widetilde{\pi}^*(x)}\ ,
	\end{align}
\end{subequations}
Solution to these equations with boundary conditions \eqref{tilpi,tilgamma-BCs+} is
\begin{subequations}\label{Formfactors}
	\begin{align}
		\label{PiPi*Formfac_}
		& \Av{|\widetilde{\pi}(x)|^2}= \mathcal{D}\left\{ \exp \left[ \frac{2}{L_{b}}\left( L/2-x\right) \right] +\Gamma
		\right\}\ ,\\
		\label{GaGa*Formfac_}
		& \Av{|\widetilde{\gamma}(x)|^2}= \mathcal{D}\left\{ \exp \left[ \frac{2}{L_{b}}\left( L/2-x\right) \right] -\Gamma
		\right\}\ ,\\
		\label{PiGa*Formfac_}
		& \Av{\widetilde{\pi}(x)\widetilde{\gamma}^*(x)}= \,i\frac{|t_+|^2}{4}\left|\frac{k_{spp}}{\varkappa'_{s}} +1\right|^2
		\mathcal{R}_s^*e^{-i\varkappa'_{s}L} \exp \left[- \frac{1}{L_{d}}\left( L/2-x\right) \right]\ ,\\[6pt]
		\label{GaPi*Formfac_}
		& \Av{\widetilde{\gamma}(x)\widetilde{\pi}^*(x)}= -i\frac{|t_+|^2}{4}\left|\frac{k_{spp}}{\varkappa'_{s}} +1\right|^2
		\mathcal{R}_se^{i\varkappa'_{s}L} \exp \left[- \frac{1}{L_{d}}\left( L/2-x\right) \right]\ ,
	\end{align}
\end{subequations}
where the notations are introduced for brevity
\begin{subequations}\label{N_Gamma-nots}
	\begin{eqnarray}
		\label{N-not}
		\mathcal{D} &=&\frac{|t_{+}|^{2}}{8}\left\vert \frac{k_{spp}}{\varkappa^{'}_{s}}+1\right\vert ^{2} e^{\varkappa _{s}''L}\left(
		1+\left\vert \mathcal{R}_{s}\right\vert ^{2}e^{-2\varkappa _{s}''L}\right),  \\
		\label{Gamma-not}
        \Gamma  &=&\frac{1-\left\vert \mathcal{R}_{s}\right\vert ^{2}e^{-2\varkappa _{s}''L}}{1+\left\vert \mathcal{R}_{s}\right\vert
			^{2}e^{-2\varkappa _{s}^{''}L}},  \\
        \label{Ld-not}
        \frac{1}{L_{d}} &=& \frac{2}{L_{f}}-\frac{1}{L_{b}}\ .
	\end{eqnarray}
\end{subequations}
Note that some terms in Eqs. \eqref{Formfactors} and \eqref{N_Gamma-nots} contain reflection coefficient $\mathcal{R}_{s}$ defined in Eq.~\eqref{Euscr_R+}.

From  Eq. \eqref{varkappa_spp} it can be seen that for small impedance fluctuations, when inequality $|\Xi|^2\ll 1$ is fulfilled, the $\mathcal{R}_{s}$ absolute value is small as compared to unity and we can neglect the terms containing this factor. In this case, the expressions for the required averages are approximately given by the following ones,
\begin{subequations}\label{Formfactors--fin}
	\begin{align}
		\label{PiPi*Formfac--fin}
		& \Av{|\widetilde{\pi}(x)|^2} \approx \frac{|t_+|^2}{2} e^{\varkappa''_{s}L}
		\left\{\exp\bigg[\frac{2}{L_b}\left(L/2-x\right)\bigg]+1\right\}\ ,\\
		\label{GaGa*Formfac--fin}
		& \Av{|\widetilde{\gamma}(x)|^2} \approx \frac{|t_+|^2}{2} e^{\varkappa''_{s}L}
		\left\{\exp\bigg[\frac{2}{L_b}\left(L/2-x\right)\bigg]-1\right\}\ ,\\
		\label{PiGa*Formfac--fin}
		& \Av{\widetilde{\pi}(x)\widetilde{\gamma}^*(x)} \approx 0\ ,\\[6pt]
		\label{GaPi*Formfac--fin}
		& \Av{\widetilde{\gamma}(x)\widetilde{\pi}^*(x)} \approx 0\ .
	\end{align}
\end{subequations}
In this approximation, formula \eqref{J_simple} can be simplified to
\begin{align}\label{J_simple1}
J(q,q^{\prime }) =\widetilde{W}\bigg( \frac{q+q^{\prime }}{2} & +\varkappa
_{s}^{\prime }\bigg)\int\limits_{-L/2}^{L/2}dX e^{-i\left( q-q^{\prime
}\right) X }\left\langle |\pi (X )|^{2}\right\rangle
\notag\\
&+\widetilde{W}\left( \frac{q+q^{\prime }}{2}-\varkappa _{s}^{\prime }\right)
\int\limits_{-L/2}^{L/2}dX e^{-i\left( q-q^{\prime }\right) X
}\left\langle |\gamma (X )|^{2}\right\rangle ,
\end{align}
where, in accordance with relations \eqref{Tilde_pi_gamma}, from Eqs.~\eqref{PiPi*Formfac--fin} and \eqref{GaGa*Formfac--fin} we get
\begin{equation}\label{pi(x)^2_gamma(x)^2}
\begin{aligned}
\left\langle |\pi (x)|^{2}\right\rangle & =\frac{|t_{+}|^{2}}{2}e%
^{\varkappa _{s}^{\prime \prime }\left( L-2x\right) }\left\{ \exp \left[
\frac{2}{L_{b}}\left( L/2-x\right) \right] +1\right\} , \\
\left\langle |\gamma (x)|^{2}\right\rangle & =\frac{|t_{+}|^{2}}{2}e%
^{\varkappa _{s}^{\prime \prime }\left( L+2x\right) }\left\{ \exp \left[
\frac{2}{L_{b}}\left( L/2-x\right) \right] -1\right\} ,
\end{aligned}
\end{equation}
In what follows, we will assume that the length of the dissipative delay of the SPP field is much larger that the length $L$ of the disordered segment   ($\varkappa _{s}^{\prime \prime}L\ll 1$), so factors $e%
^{\varkappa''_{s}\left( L \pm 2x\right) }$ in \eqref{pi(x)^2_gamma(x)^2} may be put equal to unity.
%
\subsection{Scattering pattern of the surface polariton}
\label{Scatt_diagram}
\begin{figure}[h]
	\centering
	\scalebox{.6}[.6]{\includegraphics{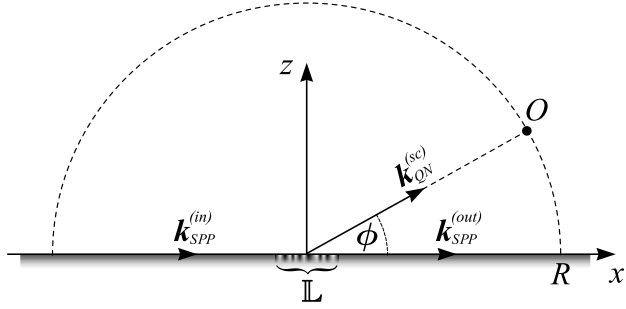}}
	\caption{Scattering scheme of a surface polariton incident on the
		surface region $\mathbb{L}$ with randomly inhomogeneous impedance, $O$ is the point of location of the receiver of scattered waves radiated into the dielectric half-space.
		\hfill\label{fig3}}
\end{figure}

\begin{figure}[h!]
	\centering
	{
		\scalebox{.9}[.9]{\includegraphics{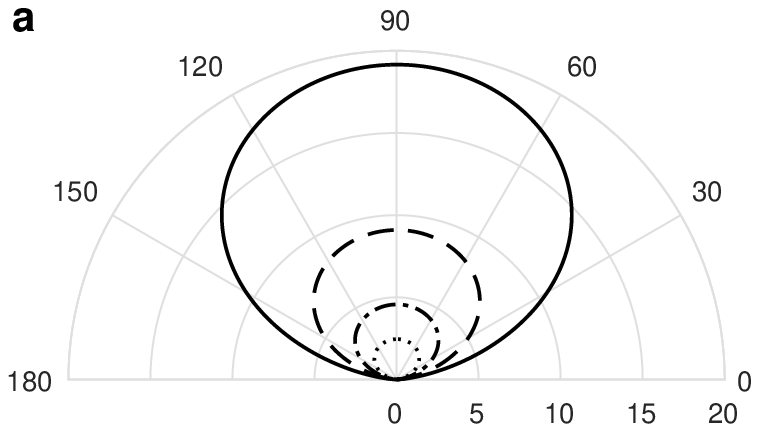}}\qquad\qquad
		\scalebox{.9}[.9]{\includegraphics{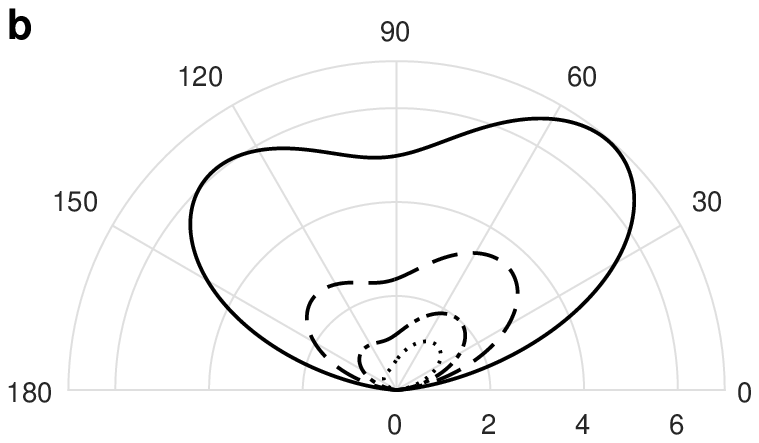}} \\[10pt]
		\scalebox{.9}[.9]{\includegraphics{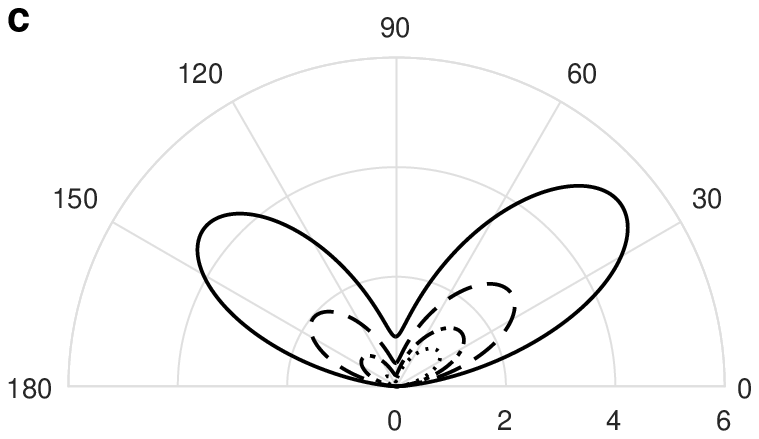}}\qquad\quad
		\scalebox{.9}[.9]{\includegraphics{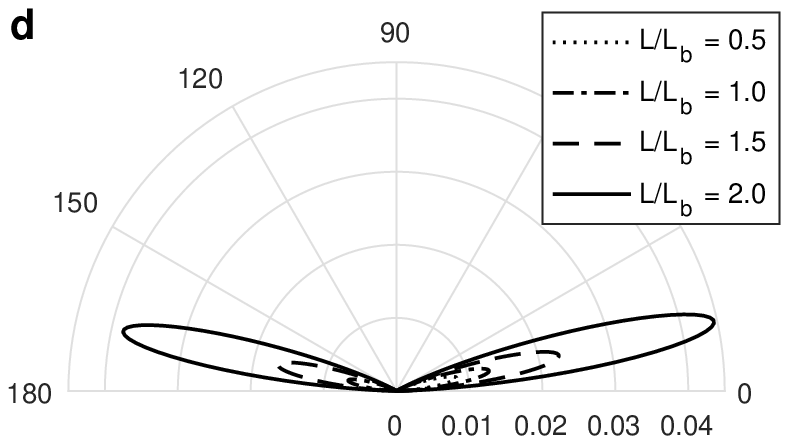}}
	}
	\caption{Polar plot of the SPP scattering by random-impedance boundary segment  $\mathbb{L}$ at different values of the parameter  $kr_c$: \textbf{\textsf{a}} -- $kr_c=0.1$; \textbf{\textsf{b}} -- $kr_c=1$; \textbf{\textsf{c}} -- $kr_c=1.5$; \textbf{\textsf{d}} -- $kr_c=10$.
		\hfill\label{fig4-new}}
\end{figure}

%
From Eqs.~\eqref{Int_q1q2_h(r)} and \eqref{J_simple1} we can obtain the scattering pattern of the SPP energy radiation induced by inhomogeneous surface segment considering it as the radiation source. For clarity we depict it as the polar plot with the center at point $(x=0,z=0)$. The average intensity at distance $R$ from this point has the form
\begin{align}\label{Intens_z>0_polar--1}
	\Av{|h(\mathbf{r})|^2} = \frac{| \bm{\Xi}|^2}{2}
	& \iint\limits_{-\infty}^{\quad\infty}\frac{dqdq'}{(2\pi)^2}
	\tfrac{\exp\big\{i(q-q')R\cos\phi+i\left[\sqrt{k^2-q^2}-\left(\sqrt{k^2-q'^2}\right)^*\right]R\sin\phi\big\}} {\left[\zeta_0+\sqrt{1-(q/k)^2}\right] \left[\zeta_0+\sqrt{1-(q'/k)^2}\right]^*}J(q,q')\ .
\end{align}
At large distances from segment $\mathbb{L}$ the radiation source can be considered as a~point one (see Fig.~\ref{fig3}).
The integrals over $q$ and $q'$ in Eq.~\eqref{Intens_z>0_polar--1} can be evaluated with asymptotic accuracy using the stationary phase method. The points of stationarity are determined from  equation
\begin{align}\label{St_phase-eq}
	R\cos\phi-X=\frac{q_{st} R\sin\phi}{\sqrt{k^2-q_{st}^2}}\ .
\end{align}
They fall on the integration axis only if $k^2-q_{st}^2>0$. Otherwise, they are saddle points located in the complex plane of the variable $q$. From Eq.~\eqref{St_phase-eq},
\begin{align}\label{St_phase-point}
	q_{st}=k\left(\cos\phi-\frac{X}{R}\sin^2\phi\right) \approx k\cos\phi\ .
\end{align}
The characteristic ``width'' of the exponent near this point is estimated as
\begin{align}\label{Exp_width}
	(\Delta q)_{ex}\sim \sqrt{\frac{k}{R}}\sin\phi\ .
\end{align}
If this width is small compared to the width of the Fourier transform of the correlation function ($\sim 1/r_c$), which is equivalent to inequality
\begin{align}\label{D_exp<<D_W}
	kr_c\frac{r_c}{R}\sin^2\phi\ll 1\ ,
\end{align}
we obtain the expression as follows for average intensity of field $h(\mathbf{r})$, viz.,
\begin{align}\label{Intens_z>0_polar--2}
	\Av{|h(\mathbf{r})|^2}
	\approx \frac{| \bm{\Xi}|^2 |t_{+}|^{2}}{4} & \frac{kL}{2\pi R}\frac{\sin^2\phi}{\big|\zeta_0+\sin\phi\big|^2} \times
\notag\\
   & \times \Bigg\{\widetilde{W}(k_{spp}+k\cos\phi)
	\left[\frac{L_b}{2L}\left(e^{2L/L_b}-1\right)+1\right] +
	\notag\\
	& \phantom{ \Bigg\{ }
    +\widetilde{W}(k_{spp}-k\cos\phi)
	\left[\frac{L_b}{2L}\left(e^{2L/L_b}-1\right) -1 \right]\Bigg\}\ .
\end{align}
This expression allows us to get answers to some questions about the radiation of the incident SPP
into the upper dielectric half-space. Fig.~\ref{fig4-new} shows the dimensionless intensity of the field radiated from segment~$\mathbb{L}$. Noteworthy is the pronounced anisotropy of the scattering pattern, which may be effectively tuned by adjusting the inhomogeneity length parameters, $r_c$ and $L$, thus serving as a tool for probing these parameters in practice.
The crucial dependence of the intensity \eqref{Intens_z>0_polar--2} on the ratio $L/L_b$ is characteristic of one-dimensional localization problems, where $L_b$ is associated with Anderson localization length \cite{LifGredPast88}. The appearance of this ratio in the context of the problem elaborated herein suggests a hidden presence of 1D localization even though the originally formulated scattering problem is not one-dimensional at the mathematical level.

Indeed, formula \eqref{Intens_z>0_polar--2} at first glance may seem to be unrelated to strictly one-dimensional wave transport. But if one returns to expression \eqref{Intens_RadField} for the average intensity it can be noticed that the related formula \eqref{|h(r)|^2-appr} may be interpreted, due to specific structure of correlator \eqref{1st_corr_orig}, as a series of transitions from one purely one-dimensional building block to another one via the CP states hidden in the operator $\hat{\mathcal{L}}$ matrix elements. By the building blocks we mean the correlators formed by averaging the pair of trial fields $\widetilde{\mathcal{H}}^{(in)}_0(q_{1,2})$, which are purely one-dimensional objects by definition. The length $L_b$, at which one-dimensional eigenstates in the random medium are known to be localized, appears as a result of the calculation of such ``two-point'' correlators in the coordinate representation.

With formula \eqref{|h(r)|^2-appr}, we are tracking actually the radiation of not a free SPP that moves along a single coordinate, but rather of 1D wave (still the same SPP) which is in the localized state formed dynamically during its multiple backscattering.  The ability of such waves to propagate along  the disordered interval of the surface depends, firstly, on the ratio between its intrinsic localization length and the size of the obstacle (segment $\mathbb{L}$ in our case), and secondly, on the efficiency of its \textit{local} transformation from a surface wave to a bulk one. As a measure of this transformation serves the operator $\hat{\mathcal{L}}$ norm, which reflects the distribution of the scattered wave energy between surface and bulk modes in the one act of such a combined scattering process.  The effect of this (single or multiple) scattering can be understood through the calculation of scattering coefficients $t_+$ and $r_-$, which have been introduced phenomenologically in Eqs.~\eqref{H_0^<} and \eqref{H_0^>}.
%
\section{The search for the scattering coefficients}
\label{Scatt_Coeffs}
%
Since there are two unknown quantities to be determined, namely, the coefficients $t_+$ and $r_-$, then, ideally, there should also be two equations for their determination. To obtain them we will use, firstly, the total flux of electromagnetic energy conservation law, and, secondly, the boundary conditions at the ends of the disordered segment of the interface.
%
\subsection{Energy flow conservation}
\label{Flow_Cons}
%
Flow channels in our problem are: 1) the incident and the reflected SPP waves~\eqref{H_0^<}, 2)~the transmitted SPP wave \eqref{H_0^>}, 3)~the radiation channel (leaking field $h(\mathbf{r})$), and 4)~the dissipation channel (losses within the metal). With the impedance description of the interface, the last channel is automatically accounted for by real part of the surface impedance  ($\zeta'_0>0$) and, correspondingly, by dissipation length \eqref{L(SPP)_dis}.

As for the first and the second channels, we will calculate the energy fluxes in them at the ``entry'' and ``exit'' points of the scattering segment $\mathbb{L}$ ($x=-L/2$ and $x=L/2$, respectively). The total energy flux in each of these channels can be calculated by integrating over $0<z<\infty$ the plasmon energy density averaged over the time, $\mathrm{w}_{spp}=\frac{1}{16\pi}\left(|\mathbf{H}_{spp}|^2+ |\mathbf{E}_{spp}|^2\right)$, multiplied by SPP phase velocity, $|\mathrm{v}_{spp}|=c/\sqrt{1+{(\zeta_0'')}^2}$, taken with the appropriate sign. In the radiation channel, the value of phase velocity does not depend on the direction of radiation, being equal to the speed of light in vacuum. The magnetic field $\mathbf{H}_{spp}$ has only one nonzero component, $H_y=H(\mathbf{r})$, and the nonzero components of the electric field, $E_x$ and $E_z$, are expressed in terms of $H(\mathbf{r})$ using Maxwell's equations in vacuum,
\begin{align}\label{ExEz->Hy}
	-ik\mathbf{E}=[\mathbf{\nabla}\times \mathbf{H}]\ \Longrightarrow
	\begin{cases}
		E_x=-(i/k)\partial H_y/\partial z\ ,\\
		E_z=(i/k)\partial H_y/\partial x\ .
	\end{cases}
\end{align}
Using \eqref{H_0^<} and \eqref{H_0^>}, the electromagnetic field energy density at the ``entry'' and ``exit'' of segment $\mathbb{L}$ can be presented in the following form,
\begin{subequations}\label{LeftRight_EnDens}
	\begin{align}
		\label{Left_EnDens}
		& \mathrm{w}_{spp}^{(\pm)}(-L/2,z)=\frac{1}{16\pi}
		\begin{pmatrix}
			1\\[3pt]|r_-|^2
		\end{pmatrix}
		\left[\big(1+|\zeta_0|^2\big)+
		\left|\sqrt{1-\zeta_0^2}\right|^2\right]e^{2\zeta''_0kz}\ ,\\
		\label{Right_EnDens}
		& \mathrm{w}_{spp}(L/2,z)=\frac{|t_+|^2}{16\pi}
		\left[\big(1+|\zeta_0|^2\big)+\left|\sqrt{1-\zeta_0^2}\right|^2\right]e^{2\zeta''_0kz}\ .
	\end{align}
\end{subequations}
The superscript $(\pm)$ in Eq.~\eqref{Left_EnDens} labels the components of the electromagnetic SPP field propagating in positive and negative directions of the $x$-axis, respectively. By integrating both densities \eqref{LeftRight_EnDens} over $z$ and multiplying them by the SPP phase velocity taken with the appropriate sign we obtain the required expressions for the energy fluxes at both ends of disordered segment~$\mathbb{L}$,
\begin{subequations}\label{LrftRight_flows}
	\begin{align}
		\label{Left(+)flow}
		& J^{(+)}(-L/2)=
		\frac{c}{32\pi k|\zeta''_0|\sqrt{1+{(\zeta_0'')}^2}}\Big[\big(1+|\zeta_0|^2\big)+
		\left|1-\zeta_0^2\right|\Big]\ ,\\
		\label{Left(-)flow}
		& J^{(-)}(-L/2)= -|r_-|^2J^{(+)}(-L/2)\ ,\\
		\label{Right(+)flow}
		& J^{(+)}(L/2)= |t_+|^2J^{(+)}(-L/2)\ .
	\end{align}
\end{subequations}
If the radiation scattering channel and the channel of dissipative loss were absent, then, equating the sum of the fluxes on the left of $\mathbb{L}$  to the flux on the right, we would obtain the equation well-known in one-dimensional scattering problems,
\begin{align}\label{1D_flow_conserv}
	|r_-|^2+|t_+|^2=1\ ,
\end{align}
which is known to express the flow conservation law.  However, in order to compose the correct balance of energy flows in the case we consider here it is necessary to add the ``radiation'' flux emitted into the free (upper) half-space.  To do this, one should take the magnetic field  \eqref{Full_solution_h(r)} at the points of the semicircle shown in Fig.~\ref{fig3}, calculate the electric field at the same points, and then, applying the formula for the time-averaged Poynting vector, calculate the total flow of electromagnetic energy through the semicircle indicated above (in fact, through the half-cylinder of unit length).

For the flow through the cylindrical surface, it is natural to calculate it in cylindric coordinate system with unit vectors $\bm{e}_r$, $\bm{e}_{\phi}$, and $\bm{e}_y$ related to cartesian unit vectors $\bm{e}_x$, $\bm{e}_y$, and $\bm{e}_z$ by equalities
\begin{equation}\label{Unit_vect_cyl-Dec}
	\begin{aligned}
		\begin{cases}
			& \bm{e}_r=\bm{e}_x\cos\phi+\bm{e}_z\sin\phi\ ,\\
			& \bm{e}_{\phi}= -\bm{e}_x\sin\phi+\bm{e}_z\cos\phi\ .\\
			& \bm{e}_y=\bm{e}_y\ ,
		\end{cases}
	\end{aligned}
\end{equation}
The radial component of the time-averaged Poynting vector has the form
\begin{align}\label{Sr->Ez_Ex}
	\overline{{S}}_r=\frac{c}{8\pi}\Re\left[\big(-E_z\cos\phi+E_x\sin\phi\big)H_y ^*\right]\ ,
\end{align}
where the bar above the symbol denotes averaging over time. Substituting here the electric field components found from Eq.~\eqref{ExEz->Hy} we get
\begin{align}\label{Sr->H_H*}
	\overline{{S}}_r = \frac{c}{8\pi k}\Im\bigg[
	\left(\frac{\partial H}{\partial x}\cos\phi+\frac{\partial H}{\partial z}\sin\phi\right)H_y ^*\bigg]\ .
\end{align}
Expression \eqref{Sr->H_H*} after averaging over realizations of random impedance can be rewritten as
\begin{align}\label{Sr=Real}
	\Av{\overline{{S}_r(\mathbf{r})}}=\frac{c}{8\pi k}
	\Im\AV{\left[\frac{\partial h(\mathbf{r})}{\partial x}\cos\phi+
		\frac{\partial h(\mathbf{r})}{\partial z}\sin\phi\right]h^*(\mathbf{r})}\ .
\end{align}
The mean flow through the lateral surface of the half-cylinder whose cross-section is schematically shown in Fig.~\ref{fig3} can be found by substituting into  Eq.~\eqref{Full_solution_h(r)} ${x=R\cos\phi}$, $z=R\sin\phi$ and then integrating the resulting expression over interval  $0<\phi<\pi$. Let us denote the radiation fluxes associated with the first and the second addenda in  Eq.~\eqref{Sr=Real} by symbols $\Av{J^{(rad)}_{1}}$ and $\Av{J^{(rad)}_{2}}$, respectively. The explicit form of the first of these terms is as follows:
\begin{align}\label{Rad_term-1}
	\Av{J^{(rad)}_{1}} & = \frac{cR}{8\pi k}\Re\int\limits_0^{\pi}d\phi\cos\phi
	\int\limits_{-\infty}^{\infty}\frac{qdq}{2\pi}\exp\left(iqR\cos\phi+i\sqrt{k^2-q^2}R\sin\phi\right) \times
	\notag\\ & \times
	\int\limits_{-\infty}^{\infty}\frac{dq'}{2\pi}\exp\left[-iq'R\cos\phi-i\Big(\sqrt{k^2-q'^2}\Big)^*R\sin\phi\right]
	\AV{\widetilde{\mathcal{R}}(q)\widetilde{\mathcal{R}}^*(q')}\ .
\end{align}
The double integral over $q$ and $q'$ in  Eq.~\eqref{Rad_term-1} coincides up to the integrand factor $q$ with the analogous integral appearing in formula  \eqref{Intens_gen}. Therefore, it can be calculated in the same way (by means of the stationary phase method) as was used earlier to obtain  formula  \eqref{Intens_z>0_polar--2}. The result of these calculations is
\begin{align}\label{Rad_term-1_fin}
	\Av{J^{(rad)}_{1}}= & |t_+|^2|\bm{\Xi}|^2 \frac{ckL}{32\pi^2} \Re\int\limits_0^{\pi}d\phi\frac{\sin^2\phi\cos^2\phi}{\big|\zeta_0+\sin\phi\big|^2}\times
	\notag\\
	& \times \Bigg\{\widetilde{W}(k\cos\phi+k_{spp})
	\left[\frac{L_b}{2L}\left(e^{2L/L_b}-1\right)+1\right]+
	\notag\\
    &\qquad\qquad +\widetilde{W}(k\cos\phi-k_{spp})
	\left[\frac{L_b}{2L}\left(e^{2L/L_b}-1\right) - \right]\Bigg\}\ .
\end{align}

The expression for the second term in formula \eqref{Sr->H_H*}, $\Av{J^{(rad)}_{2}}$, differs from that of Eq.~\eqref{Rad_term-1} only in that the integrand factor $q$ in it must be replaced with $\sqrt{k^2-q^2}$, and factor $\cos\phi$  in the integral over $\phi$ with factor $\sin\phi$. At the point of phase stationarity $\sqrt{k^2-q^2}=k\sin\phi$, and we get
\begin{align}\label{Rad_term-2_fin}
	\Av{J^{(rad)}_{2}}= & |t_+|^2|\bm{\Xi}|^2 \frac{ckL}{32\pi^2} \Re\int\limits_0^{\pi}d\phi\frac{\sin^4\phi}{\big|\zeta_0+\sin\phi\big|^2}\times
	\notag\\
	& \times \Bigg\{\widetilde{W}(k_{spp}+k\cos\phi)
	\left[\frac{L_b}{2L}\left(e^{2L/L_b}-1\right) +1 \right] +
	\notag\\
	& \phantom{+\widetilde{W}k\cos\phi}
	+\widetilde{W}(k_{spp}-k\cos\phi)
	\left[\frac{L_b}{2L}\left(e^{2L/L_b}-1\right)-1 \right]\Bigg\}\ .
\end{align}
The flow balance now takes the form
\begin{align}\label{Flow_balance}
	J^{(+)}(-L/2)+J^{(-)}(-L/2)=J^{(+)}(L/2)+\Av{J^{(rad)}}\ ,
\end{align}
which is no longer reduced to equation \eqref{1D_flow_conserv} as for strictly one-dimensional problems. Dividing Eq.~\eqref{Flow_balance} by the factor of $J^{(+)}(-L/2)$ we get the flow balance equation in the following form,
\begin{align}\label{Flow_balance-fin}
	1-|r_-|^2=|t_+|^2\left(1+\Av{I^{(rad)}}\right)\ .
\end{align}
Here, $\Av{I^{(rad)}}$ is the \textit{normalized} radiation flux, which in the limit of  $\|\hat{\mathcal{L}}\|\ll 1$ is represented by formula
\begin{align}\label{Norm_rad_flow}
	\Av{I^{(rad)}}=&  |\bm{\Xi}|^2\frac{k^2L}{\pi}
	\frac{|\zeta''_0|\sqrt{1+{(\zeta_0'')}^2}}{\big(1+|\zeta_0|^2\big)+\left|1-\zeta_0^2\right|}
	\Re\int\limits_0^{\pi}d\phi\frac{\sin^2\phi}{\big|\zeta_0+\sin\phi\big|^2} \times
	\notag\\
	& \times \Bigg\{\widetilde{W}(k_{spp}+k\cos\phi)
	\left[\frac{L_b}{2L}\left(e^{2L/L_b}-1\right)+1\right] +
    \notag\\
	&\qquad\qquad +	\widetilde{W}(k_{spp}-k\cos\phi)
	\left[\frac{L_b}{2L}\left(e^{2L/L_b}-1\right)-1\right]\Bigg\}\ .
\end{align}
%
\subsubsection{On the possibility to control the radiation flux by polar angle $\phi$}
\label{I(rad)-anisotropy}
%
Functions $\widetilde{W}(k_{spp}\pm k\cos\phi)$ in expressions \eqref{Intens_z>0_polar--2} and \eqref{Norm_rad_flow} may be either smooth or sharp in the range of angle $\phi$, depending on the relationship between the correlation radius ($r_c$) and the SPP wavelength ($\lambda_{spp}\sim k^{-1}_{spp}\sim k^{-1}$). If $kr_c\ll 1$, both of these functions differ only slightly from $\widetilde{W}(0)$, and the angular dependence of the energy flux is mainly determined  by factor
\begin{align}\label{Phi_distrib-smooth}
	\mathcal{E}(\phi)=\frac{\sin^2\phi}{\big|\zeta_0+\sin\phi\big|^2}\ .
\end{align}
Function $W(x)$ depends actually on \textit{dimensionless argument} $x/r_c$. Hence, if $kr_c\gg 1$, then its Fourier components $\widetilde{W}(k_{spp}\pm k\cos\phi)$ in Eq.~\eqref{Intens_z>0_polar--2} are small as compared to the value of $\widetilde{W}(0)$, because at almost any angle $\phi$  the estimate is well-grounded $|k_{spp}\pm k\cos\phi|\sim k$. So, it appears that it is practically impossible to significantly influence the \textit{value} of the energy flux in the radiation field. The angular dependence only can be noticeably influenced,  and the main controlling parameter here is the ratio of the correlation radius, which specifies the width of the correlation functions in Eqs.~\eqref{Intens_z>0_polar--2} and \eqref{Norm_rad_flow}, to the SPP wavelength, i.\,e. the parameter $kr_c$.
%
\subsection{Boundary conditions at the ends of the disordered surface segment}
\label{End_BCs}
%
The flow balance equation \eqref{Flow_balance-fin}, which relates coefficients $r_-$ and $t_+$ to each other, is not sufficient to determine these coefficients rigorously. To find one more constraint equation we refer to boundary conditions at the contact points $x=\pm L/2$ of segment $\mathbb{L}$ with the ``outer'' regions of the interface corresponding to $|x|>L/2$. According to our model, the total magnetic field of the wave is sought as the sum of ``seed'' field $H_0(\mathbf{r})$ and the \textit{leaking} (radiation) field $h(\mathbf{r})$. The latter vanishes at the ends of the segment, and therefore the ``outer'' (with respect to~$\mathbb{L}$) SPP fields and the ``internal'' seed field $H_0^{(int)}(\mathbf{r})$ declared in the form \eqref{H_0^in} must be binded at these points together. The binding should be done solely at $z=0$ because on vertical lines $x=\pm L/2$ at $z>0$ the field  $h(\mathbf{r})$ in general is not equal to zero. In other words, at the ends of  $\mathbb{L}$ we must bind the fields \eqref{H_0^<} and \eqref{H_0^>} from the outside of the segment with the field \eqref{H_0^in} inside it.

We sought the field inside $\mathbb{L}$ in the form \eqref{H^in->pi_gamma}, and the BCs \eqref{tilpi,tilgamma-BCs+} and \eqref{tilpi,tilgamma,G-BC-} for functions $\pi(x)$ and $\gamma(x)$ were obtained. Both of these BC sets were obtained by fitting the internal seed field with external SPP fields at $z=0$. We could relate the BCs at different ends of  segment $\mathbb{L}$ proceeding directly from the dynamic equations, thus obtaining the relationship between the reflection and the transmission coefficient. Yet, for this it would be necessary for us to solve equations~\eqref{pi_gamma-eqs-new} exactly, which goal cannot be reached in view of the random nature of functions $\eta(x)$ and $\xi_{\pm}(x)$.

Nevertheless, we can establish the connection between two-side BCs without solving directly the dynamic equations, yet with statistical accuracy, within the so-called ``correlation approximation''. Applying the averaging procedure described in section \ref{Scatt_technique} to equations \eqref{pi_gamma-eqs-new2} we get the following result,
\begin{subequations}\label{<pi-><gamma->-fin}
	\begin{align}
		\label{<pi->-fin}
		& \Av{\widetilde{\pi}(x)}=\frac{t_+}{2}\left(\frac{k_{spp}}{\varkappa'_{s}}+1\right)
		e^{-i\varkappa_{s}L/2}
		\exp\Bigg[-\frac{1}{2}\left(\frac{1}{L_f} -\frac{1}{L_b}\right) \big(L/2-x\big)\Bigg]\ ,\\
		\label{<gamma->-fin}
		& \Av{\widetilde{\gamma}(x)}=\frac{t_+}{2i}\left(\frac{k_{spp}}{\varkappa'_{s}}-1\right)
		e^{i\varkappa_{s}L/2}
		\exp\Bigg[-\frac{1}{2}\left(\frac{1}{L_f} -\frac{1}{L_b}\right) \big(L/2-x\big)\Bigg]\ .
	\end{align}
\end{subequations}
Equalities \eqref{<pi-><gamma->-fin} allow to relate the values of mean (not exact!) field $\Av{\mathcal{H}^{(int)}_0(x)}$ on the inner sides of both ends of segment $\mathbb{L}$ and then, by fitting them with ``external'' fields \eqref{H_0^<} and \eqref{H_0^>}, to establish the connection between parameters $r_-$ and $t_+$.

The fitting at $x=L/2$ with the use of Eqs.~\eqref{<pi-><gamma->-fin} proceeds automatically because the right-hand BC \eqref{tilpi,tilgamma-BCs+} has already been used when performing the statistical averaging. Now, with the same functions \eqref{<pi-><gamma->-fin} we can satisfy the fitting condition at $x=-L/2$, which results in equation as follows,
\begin{align}\label{Left_BC}
	1+r_-=\frac{t_+}{2}\exp\left[-\left(\frac{1}{L_f}-\frac{1}{L_b}\right)\frac{L}{2}\right]
	\left[\left(\frac{k_{spp}}{\varkappa'_{s}} +1\right)e^{-i\varkappa_{s}L}-
	\left(\frac{k_{spp}}{\varkappa'_{s}}-1\right)e^{i\varkappa_{s}L}\right]\ .
\end{align}
Note some peculiar features of Eq. \eqref{Left_BC}. First of all, for $|\Xi|^2\ll 1$ the second term in square brackets of the rightmost factor is small in magnitude in comparison with the first one, and thus it can be neglected. In this case, the factor at the exponential function in the first term can be, with good accuracy, replaced with the value of 2.

The second feature relates to the exponential factor containing the difference between the reciprocal scattering lengths $L_{f}$ and $L_{b}$. The absolute value of quantity $\left(1/L_f - 1/L_b\right)L/2$ may become small as compared to unity in two rather specific cases. The first is the so-called \textit{ballistic limit} corresponding to inequality $L\ll L_{f,b}$. The other occasion appears when the difference between the indicated reciprocal scattering lengths is made small in comparison with~$L^{-1}$. Such a possibility is quite realizable, as can be seen from \eqref{1/Lf-1} and \eqref{1/Lb-1},  if correlation radius $r_c$ is small as compared to the wavelength (i.\,e., if $kr_c\ll 1$).

Regardless of these two specific cases, in the limit of $|\Xi|^2\ll 1$ equation \eqref{Left_BC} reduces to the shortened form
\begin{equation}\label{1+r-=t_+exp}
	1+r_-\approx t_+\exp\left[-\left(\frac{1}{L_f}-\frac{1}{L_b}\right)\frac{L}{2}\right]e^{-i\varkappa_{s}L}\ .
\end{equation}
Suppose, that the seed surface wave is Anderson-localized ($L_b\ll L$) within segment $\mathbb{L}$. This would mean that if it hits the obstacle on the left it can penetrate into it no more than for a distance $\sim L_b$. Then the right-hand part of Eq.~\eqref{1+r-=t_+exp} becomes negligibly small, and we arrive asymptotically at the equality
\begin{equation}\label{r=-1}
	r_-\approx -1\ ,
\end{equation}
which implies that the incident SPP is almost completely reflected from the imperfect boundary region in the backward direction.  Moreover, as follows from balance equation \eqref{Flow_balance-fin}, with the increase in length $L$, the transmission coefficient tends to zero \textit{exponentially fast} in order to compensate exponentially large factors $\exp(2L/L_b)$ in the right-hand part of expression \eqref{Norm_rad_flow} for the mean radiant flux.

If the seed SPP is not exponentially localized within $\mathbb{L}$ ($L_b\gtrsim L$), i.\,e., it propagates within that segment ($L_f\gtrless L$) ballistically or diffusively, the equality~\eqref{r=-1} is violated. As a result, the radiation pattern to a great extent loses its directionality and approaches the form shown in Fig.~\ref{fig4-new}a. In this case, the total value of the radiation flux is not necessarily small, since all the results discussed above refer to the case of small operator $\hat{\mathcal{L}}$ norm, which implies strong intermixing of surface and bulk scattering modes.
%
\section{Discussion of the results and prospects}
%
To conclude, in this paper we have constructed a detailed theory of surface plasmon--polariton scattering by a finite region of metal-dielectric interface with random surface impedance. To assess the degree of coupling between bulk and surface scattered modes we suggested and justified the appropriate criterion, viz., the value of Hilbert norm of the operator describing the intermediate scattering states we entitle \textit{composite polaritons}. The term ``composite'' refers to the mixed states which represent the weighted sum of truly \emph{surface} polaritons and specific \textit{bulk} states we refer to as quasi-Norton waves, which are responsible for energy leakage from the surface into the free dielectric half-space.

The waves transferring plasmon--polariton energy into the bulk of the dielectric have attracted considerable attention in recent years. In particular, in Ref.~\cite{SarShubShal99} the attempt was made (in our view, not
quite reasonable) to explain the experimentally observed sharp peaks of the local field in random metal-dielectric composites by Anderson localization of surface polaritons. In Ref.~\cite{StockmanFaleevBergman01} such an interpretation of the peaks
was rejected and, instead of it, the concept of so-called ``dark local modes'' that arise in the ``near zone'' only of the sources and cannot be excited from the ``far zone'' was proposed.

In Ref.~\cite{LalHug06}, the near-field oscillations at the inhomogeneous interface between two media with different dielectric constants were called ``creeping waves'', in contrast to the ``evanescent-type excitations'' so named in Ref.~\cite{Gay06} where these waves, apparently for the first time, were discovered experimentally. The relationship between creeping waves and the integrals over cut edges that arise in our theory (see section \ref{Full_Sol} of the present work), was noted in Ref.~\cite{GravelSheng08}, where these waves were given a different name, viz., ``transient plasmon--polaritons''. We believe, however, that the closest correspondence of waves characterized by propagator \eqref{Int_G(SPP)_branch} to local field effects observed in experiments was noted in Refs.~\cite{Nikitin09,Nikitin10,Nikitin11}, where these waves were associated with Norton waves which are emitted by a dipole source located near the earth's surface \cite{Norton36,Norton37}.

Regardless of the specific name, the common thing for all the above cited papers is the fact that they discussed
the \emph{near fields} excited by quasi-point sources located on a \emph{uniform} impedance surface. The issues related to this field were elaborated, though mainly experimentally, also in paper~\cite{Many-many08}. In the present work, we make use of the term ``quasi-Norton waves'' for the harmonics described by the cut-edge term in the integral \eqref{G(SPP)(x,x')} in view of the closest, in our view, correspondence of their mathematical structure just to Norton waves.

Under conditions of weak coupling between surface and bulk scattered harmonics, which is possible only at a sufficiently high level of dissipation in the conductor, the main result of the scattering is the emission of some part of the incident SPP energy from the surface of the conductor into the bulk of the dielectric/vacuum. We managed to calculate the radiation pattern and show that the radiated energy is proportional to the Fourier transform $\widetilde{W}(q)$ of the two-point correlation function of the impedance taken at points ${q_{\pm}=k_{spp}\pm k\cos\phi}$. This dependence has obvious similarities with angular dependence of the field scattered by periodic reflecting gratings, for which Wood~\cite{Wood1902} has discovered the anomalies in the form of well-marked reflection resonances, the origin of which was explained later on by Fano \cite{Fano41}. In our theory, the
maxima of correlation function $\widetilde{W}(q)$ should play the role of resonances of this kind in the scattering of SPP by the \emph{random-impedance} region of the interface. Yet, for random processes of a general type such maxima are usually located near $q=0$, so the wave number differences present in our formulas usually do not fit into those regions.

Nevertheless, it is possible to achieve resonance radiation in a given direction for random-impedance gratings if we manage to generate a random process (variable part of the impedance) by presetting the required form of the correlation function. A feasible recipe for such a ``construction'' of random processes was suggested in Refs.~\cite{usatenko2009random,Maystrenko_2013}. Thus, by furnishing the impedance with the required correlation properties we can implement in practice the controlled directed radiation of the field produced by a surface plasmon--polariton in the course of its scattering by the defects of the interface.

For the parameters of the impedance and the plasmon--polariton wave such that the coupling of surface and bulk scattered harmonics is not weak but, rather, is strong enough the main effect of such scattering reduces to almost mirror reflection of the incident SPP and the suppression of quasi-Norton component of its radiation. Note that the strong SPP scattering can occur not only due to the large dispersion of the impedance but also due to the decrease in the dissipative losses in the conductor. The norm of the operator describing the mixing of scattered modes is inversely proportional to the dissipative part of the impedance and can be made arbitrarily large even for small yet finite values of its reactive component.

This situation is similar to the one occuring when a normally incident plane wave is entirely (with no dissipation taken into account) back-reflected by a semi-finite one-dimensionally disordered medium due to Anderson localization of wave states in it \cite{LifGredPast88}. The difference between our problem and the aforementioned problem of purely one-dimensional scattering is that the propagation medium in our case is not one-dimensional from the very beginning. Moreover, the system we study in view of its total openness is pronouncedly non-Hermitian. In non-Hermitian systems the Anderson localization is, as a rule, significantly suppressed. Its effectively one-dimensional character  and, as a consequence, the absence of leaking waves result from the interference of trial harmonics multiply back-scattered by the impedance fluctuations. The fact that the efficiency of the interference increases with the decrease in the dissipation rate and so does the degree of localization seems to be pretty natural.

The effective regulation of the SPP energy leakage into the bulk of the dielectric at strong coupling between bulk and surface scattered modes is undoubtedly of great importance for intensively developing plasmonics. The artificial ``mirroring'' the obstacles that emerge along the pathway of surface wave propagation opens up the prospect to control effectively the direction of its transmission and thus to develop controllable open-type surface waveguides.
%
\appendix
%
\section{Evaluation of the operator $\hat{\mathcal{L}}$ norm}
\label{App_A}
%
It is natural to characterize the ``magnitude'' of random operator $\hat{\mathcal{L}}$ by its mean-square Euclidean norm, defined in terms of the scalar product,
\begin{equation}\label{Norm-def}
	\Av{\|\hat{\mathcal{L}}\|^2}=\sup_{\varphi}
	\frac{\Av{\big(\hat{\mathcal{L}}\varphi,\hat{\mathcal{L}}\varphi\big)}}{\big(\varphi,\varphi\big)}\ .
\end{equation}
Here $\varphi$ is the complete set of  functions on which the action of operator $\hat{\mathcal{L}}$ is defined.

Let us compare the action of operator $\hat{\mathcal{L}}=\hat{G}^{(CP)}\hat{\zeta}_L$ with the action of the unit operator on the same arbitrary vector $\varphi(x)$ of Banach space $\mathbb{B}\{\varphi\}$ whose properties we must determine. It is most convenient to work with operator $\hat{\mathcal{L}}$ in momentum representation, since in this case matrix elements of this operator have a relatively simple form (see formulas ~\eqref{Oper_notations}). The action of the operator $\hat{\mathcal{L}}$  on arbitrary function $\widetilde{\varphi}(q)\in\mathbb{B}$, due to the finiteness of function $\zeta(x)$ support, is defined by equations
\begin{equation}\label{L->phi(x)}
	\big[\hat{\mathcal{L}}\varphi\big](q)=
	\int\limits_{-\infty}^{\infty}\frac{dq'}{2\pi}
	\mathcal{L}(q,q')\widetilde{\varphi}(q')
	=\int_{\mathbb{L}}dx\mathcal{L}(q,x)\varphi(x)\ .
\end{equation}
We will be interested in trial functions that can be expanded into Fourier integral on segment $\mathbb{L}$, but not necessarily having the predetermined values at the ends of that segment. We impose the latter condition in order deal with \textit{open} segments on which the spectrum of the momentum $x$-component is not quantized.

Let us choose as a trial function an \textit{arbitrary}  function of the form  $\varphi(x)=\theta(L/2-|x|)g(x)$ which is expandable into Fourier integral. The action of an arbitrary operator $\hat{\mathcal{A}}$ on this function is defined by equality
\begin{align}\label{hat_A->phi}
	\big[\hat{\mathcal{A}}\varphi\big](x)=\int_{\mathbb{L}}dx'\bra{x}\hat{\mathcal{A}}\ket{x'}\varphi(x')=
	\int\limits_{-\infty}^{\infty}\frac{d\kappa}{2\pi}\widetilde{\varphi}(\kappa)
	\int_{\mathbb{L}}dx'\bra{x}\hat{\mathcal{A}}\ket{x'}e^{i\kappa x'}\ .
\end{align}
Moreover, matrix elements of the operator  $\hat{\mathcal{A}}$ are not necessarily concentrated within segment $\mathbb{L}$.

We define the square norm of vector $\hat{\mathcal{A}}\varphi$ in a standard way, through the scalar product,
\begin{align}\label{|A|^2}
	\|\hat{\mathcal{A}}\varphi\|^2= & \big(\hat{\mathcal{A}}\varphi,\hat{\mathcal{A}}\varphi\big)=
	\iint\limits_{-\infty}^{\quad\infty}\frac{d\kappa_1d\kappa_2}{(2\pi)^2}
	\widetilde{\varphi}^*(\kappa_1)\widetilde{\varphi}(\kappa_2) \times
\notag\\
&\times	\int\limits_{-\infty}^{\infty}dx\iint_{\mathbb{L}}dx_1dx_2
	\bra{x}\hat{\mathcal{A}}\ket{x_1}^*\bra{x}\hat{\mathcal{A}}\ket{x_2}
	e^{-i\kappa_1x_1+i\kappa_2x_2}\ .
\end{align}
If $\hat{\mathcal{A}}$ would be the unit operator on the entire $x$-axis then its matrix element on this axis was $\delta(x-x')$. Taking into account the finiteness of the support of the function expanded into the Fourier integral, the action of such an operator onto the exponent function in the right-hand part of \eqref{hat_A->phi} is determined by equality
\begin{align}\label{^1->e^(-ikappa_x)}
	\big[\hat{1} \cdot e^{i\kappa x}\theta(L/2-|x|)\big](x)= e^{i\kappa x}
	\theta(L/2-|x|)\ .
\end{align}
Based on this representation, the norm squared of the vector $\hat{1}\cdot\varphi$ equals
\begin{align}\label{|1|^2}
	\|\hat{1}\cdot\varphi\|^2=
	\iint\limits_{-\infty}^{\quad\infty}\frac{d\kappa_1d\kappa_2}{(2\pi)^2}
	\widetilde{\varphi}^*(\kappa_1)\widetilde{\varphi}(\kappa_2)\Delta_L(\kappa_1-\kappa_2)\ ,
\end{align}
where function  $\Delta_L(\kappa)$ defined in Eq.~\eqref{Underlimit_delta} has the meaning of a prelimit $\delta$-function of characteristic width~${\sim 1/L}$.

If we take as operator  $\hat{\mathcal{A}}$ the operator $\hat{\mathcal{L}}$ defined by matrix elements \eqref{Kernel_Fredholm_eq}, the action of this operator on the oscillating exponent with support  $\mathbb{L}$ in Eq.~\eqref{hat_A->phi} is defined by formula
\begin{align}\label{^L->e^(iqx)}
	\big[\hat{\mathcal{L}} \cdot e^{i\kappa_j x'}\big](\kappa)=
	\int\limits_{-\infty}^{\infty}\frac{d\kappa'}{2\pi}
	\bra{\kappa}\hat{G}^{(CP)} & \ket{\kappa'}\int_{\mathbb{L}}dx'\bra{\kappa'}
	\hat{\zeta}_L\ket{x'}e^{i\kappa_jx'}=
\notag\\
   & =\left[\zeta_0+\sqrt{1-(\kappa/k)^2}\right]^{-1}\widetilde{\zeta}(\kappa-\kappa_j)\ ,
\end{align}
the validity of which is easy to verify by calculating first the cross matrix element of the operator $\hat{\zeta}_L$. The norm squared of vector  $\hat{\mathcal{L}}\varphi$ has, thus, the form
\begin{align}\label{|L|^2}
	\|\hat{\mathcal{L}}\varphi\|^2=\iint\limits_{-\infty}^{\quad\infty} & \frac{d\kappa_1d\kappa_2}{(2\pi)^2}
	\widetilde{\varphi}^*(\kappa_1)\widetilde{\varphi}(\kappa_2) \times
\notag\\
  & \times\int\limits_{-\infty}^{\infty}\frac{d\kappa}{2\pi}
	\left|\zeta_0+\sqrt{1-(\kappa/k)^2}\right|^{-2}
	\widetilde{\zeta}^*(\kappa-\kappa_1)\widetilde{\zeta}(\kappa-\kappa_2)\ ,
\end{align}
and its average can be calculated using equality
\begin{align}
	\label{zeta*(q)_zeta(q')-corr}
	& \Av{{\widetilde{\zeta}}^*(q){\widetilde{\zeta}}(q')} = |\bm{\Xi}|^2
	\int\limits_{-\infty}^{\infty}\frac{ds}{2\pi}\widetilde{W}(s)
	\Delta_L(q^*-s)\Delta_L(s-q')\ ,
\end{align}
which is equivalent to Eq.~\eqref{zeta_zeta*-corr(x)}, yet in momentum representation. After averaging we get
\begin{align}\label{<|L|^2>}
	\Av{\|\hat{\mathcal{L}}\varphi\|^2}=|\bm{\Xi}|^2
	\iint\limits_{-\infty}^{\quad\infty} & \frac{d\kappa_1d\kappa_2}{(2\pi)^2}
	\widetilde{\varphi}^*(\kappa_1)\widetilde{\varphi}(\kappa_2) \int\limits_{-\infty}^{\infty}\frac{d\kappa}{2\pi}
	\left|\zeta_0+\sqrt{1-(\kappa/k)^2}\right|^{-2} \times \notag\\
	& \times
	\int\limits_{-\infty}^{\infty}\frac{ds}{2\pi}\widetilde{W}(s)
	\Delta_L(s-\kappa+\kappa_1)\Delta_L(s-\kappa+\kappa_2)\ . \
\end{align}
If  inequality \eqref{r_c<<L} holds true, the $\Delta_L$-functions in Eq.~\eqref{<|L|^2>} are more sharp than function $\widetilde{W}(s)$.The latter can be taken out of the integral  at the point, for example, $s=\kappa-\kappa_1$, and we get
\begin{align}\label{<|L|^2>-appr_1}
	\Av{\|\hat{\mathcal{L}}\varphi\|^2}\approx|\bm{\Xi}|^2
	\iint\limits_{-\infty}^{\quad\infty} & \frac{d\kappa_1d\kappa_2}{(2\pi)^2}
	\widetilde{\varphi}^*(\kappa_1)\widetilde{\varphi}(\kappa_2)\Delta_L(\kappa_1-\kappa_2) \times
\notag\\
   & \times
	\int\limits_{-\infty}^{\infty}\frac{d\kappa}{2\pi}
	\left|\zeta_0+\sqrt{1-(\kappa/k)^2}\right|^{-2}
	\widetilde{W}(\kappa-\kappa_1)\ .
\end{align}

\begin{figure}[h!]
	\centering
	\scalebox{.7}[.7]{\includegraphics{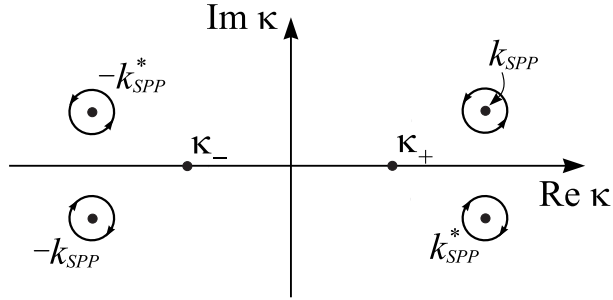}}
	\caption{Singularity points in expression~\eqref{L_norm-def-1} for the norm squared.
		\hfill\label{fig2-new}}
\end{figure}

\begin{figure}[h!]
	\centering
	\scalebox{.7}[.7]{\includegraphics{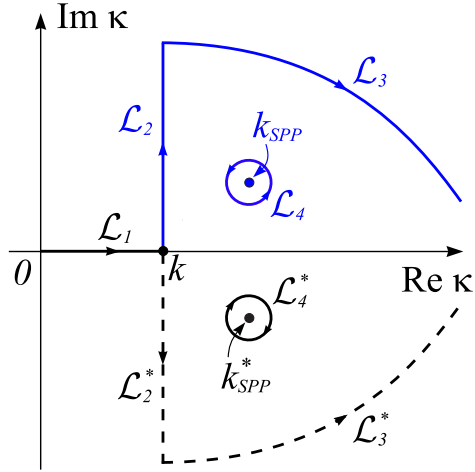}}
	\caption{Integration contour used to calculate the integral over $\kappa$ in  formula~\eqref{L_norm-def-1}.
		\hfill\label{fig3-new}}
\end{figure}

Let us now define in more detail the class of functions included in the norm definition \eqref{Norm-def}. First of all, they belong to the functional space~$L_2$, i.\,e. they are square-integrable functions since they are defined on a~finite interval and (by assumption) have no non-integrable singularities in it. In addition, we will assume that the characteristic spatial scale of the variation of these functions is either small in comparison with the length $L$ of the irregular segment or does not substantially exceed it. The latter property, however, has already been declared when introducing the factor $\theta(L/2-|x|)$ in the definition of the trial function. This implies that the characteristic scale of change of Fourier transforms of the trial functions is either large as compared with $\Delta_L$ function width or is of the same order of magnitude. The integrals over $\kappa_2$ in \eqref{|1|^2} and \eqref{<|L|^2>-appr_1} are calculated with asymptotic accuracy, and we get
\begin{subequations}\label{ApprNorms_1,L}
	\begin{align}
		\label{|1|^2-2}
		\|\hat{1}\cdot\varphi\|^2 &\approx
		\int\limits_{-\infty}^{\infty}\frac{d\kappa}{2\pi}
		\big|\widetilde{\varphi}(\kappa)\big|^2\ ,\\
		\label{<|L|^2>-appr-2}
		\Av{\|\hat{\mathcal{L}}\varphi\|^2} &\approx |\bm{\Xi}|^2
		\int\limits_{-\infty}^{\infty}\frac{d\kappa}{2\pi}\left|\zeta_0+
		\sqrt{1-(\kappa/k)^2}\right|^{-2}
		\int\limits_{-\infty}^{\infty}\frac{d\kappa_1}{2\pi}\big|
		\widetilde{\varphi}(\kappa_1)\big|^2
		\widetilde{W}(\kappa-\kappa_1)\ .
	\end{align}
\end{subequations}
Consider the integral over $\kappa_1$ in  expression \eqref{<|L|^2>-appr-2}. Both of the integrands, $\big|\widetilde{\varphi}(\kappa_1)\big|^2$ and ${\widetilde{W}(\kappa-\kappa_1)}$, have specific characteristic scales of change. For function  $\big|\widetilde{\varphi}(\kappa_1)\big|^2$ such scales can be $1/L$, $1/L^{(spp)}_{dis}$ and $1/L^{\text{(loc)}}$, where by $L^{\text{(loc)}}$, we mean the length of the \emph{one-dimensional} Anderson localization which is coincident, in order of magnitude, with the extinction length of the wave function. Taking  account of inequality \eqref{r_c<<L}, all these scales are small as compared to the inverse correlation length, $1/r_c$.  This implies that the integral over $\kappa_1$ in Eq.~\eqref{<|L|^2>-appr-2} can be calculated approximately by taking the correlation function $\widetilde{W}$ out of it at point $\kappa_1=0$.  After the division of \eqref{<|L|^2>-appr-2} by \eqref{|1|^2-2} we actually get the expression for the mean square norm of operator $\hat{\mathcal{L}}$ which does not depend on the specific structure of trial functions with the exception of qualitative assumptions we made above regarding the scale of their change and/or localization,
\begin{align}\label{L_norm-def-1}
	\Av{\|\hat{\mathcal{L}}\|^2}= |\bm{\Xi}|^2
	\int\limits_{-\infty}^{\infty}\frac{d\kappa}{2\pi}
	\left|\zeta_0+\sqrt{1-(\kappa/k)^2}\right|^{-2}
	\widetilde{W}(\kappa)\ .
\end{align}
The integral in \eqref{L_norm-def-1} can be estimated by taking into account the peculiarities of the plasmon--polariton dispersion law which corresponds to the zero of the integrand denominator. All singularity points of the denominator are depicted in Fig.~\ref{fig2-new}. These are four poles at points  $\kappa=\pm k_{spp}$ and $\kappa=\pm k^*_{spp}$, and two branch points $\kappa_{\pm}=\pm k$. The integral over $\kappa$ can be calculated by substituting into \eqref{L_norm-def-1} the correlation function $\widetilde{W}(\kappa)$ in the form of Fourier integral,
\begin{align}\label{W(kappa)}
	\widetilde{W}(\kappa)= & \int\limits_{-L}^{L}dX\, W(|X|)
	e^{-i\kappa X} = \int\limits_{0}^{L}dX\, W(X)
	\left(e^{i\kappa X}+e^{-i\kappa X}\right)=
	\widetilde{W}_+(\kappa)+\widetilde{W}_-(\kappa)\ .
\end{align}
For even function $W(x)$ formula \eqref{L_norm-def-1}, taking account of \eqref{W(kappa)}, can be transformed to the form
\begin{align}\label{L_norm-def-2}
	\Av{\|\hat{\mathcal{L}}\|^2}= 2|\bm{\Xi}|^2
	\int\limits_{0}^{\infty}\frac{d\kappa}{2\pi}
	\left|\zeta_0+\sqrt{1-(\kappa/k)^2}\right|^{-2}
	\left[\widetilde{W}_+(\kappa)+\widetilde{W}_-(\kappa)\right]\ .
\end{align}
When calculating the ``plus'' and ``minus'' terms we divide the integration contour into the ``left'' and the ``right'' intervals, see ~ Fig.~\ref{fig3-new}. The ``left'' interval,  segment $\mathcal{L}_1=[0,k]$, is common for
both terms in \eqref{L_norm-def-2}, while the ``right'' contour, the ray $[k,\infty)$ for the ``plus'' term, is represented as a sum of semi-axis $\mathcal{L}_2=[k,k+i\infty)$ and quarter-circle $\mathcal{L}_3$ of infinite radius. For the term with the minus index in \eqref{L_norm-def-2}, the ray $[k,\infty)$ is replaced by the sum of the lower semi-axis $\mathcal{L}_2^*$ and the lower arc $\mathcal{L}_3^*$.

The contribution of the integral over segment $\mathcal{L}_1$ to \eqref{L_norm-def-2} is given by
\begin{align}\label{Square_norm_L1}
	\Av{\|\hat{\mathcal{L}}\|^2}_{[0,k]}=
	4|\bm{\Xi}|^2\int\limits_{0}^{L}dX\, W(X) \int\limits_{0}^{k}\frac{d\kappa}{2\pi}
	\frac{\cos(\kappa X)}{\left|\zeta_0+\sqrt{1-(\kappa/k)^2}\right|^{2}}\ .
\end{align}
Since we are only interested in the asymptotic estimate of the norm, there is no need to calculate this integral exactly. The region of variable~$X$ which provides the main contribution to the integral is determined by function $W(X)$ and is thus concentrated at $X\sim r_c$. If $kr_c\ll 1$, then we get the following estimate for \eqref{Square_norm_L1}, viz.,
\begin{align}\label{Square_norm_L1<<}
	\Av{\|\hat{\mathcal{L}}\|^2}_{[0,k]}\sim \frac{|\Xi|^2}{|\zeta_0|^2}kr_c\ .
\end{align}
If $kr_c\gg 1$, the main contribution to the integral over $\kappa$ comes from small areas close to ends of the interval $\kappa\in[0,k]$. This results in a decrease of the integral, and thus of the whole expression \eqref{Square_norm_L1}, by factor  $\sim (kr_c)^{-1}\ll 1$. Finally, the overall estimate for \eqref{Square_norm_L1} can be written as a single formula,
\begin{align}\label{Square_norm_L1><}
	\Av{\|\hat{\mathcal{L}}\|^2}_{[0,k]}\sim \frac{|\Xi|^2}{|\zeta_0|^2} \min(1,kr_c)\ .
\end{align}

Next, let us estimate the contribution from the integrals over segments $\mathcal{L}_2$ and $\mathcal{L}^*_2$ on the contours depicted in Fig.~{\ref{fig3-new}}. The contribution of segment $\mathcal{L}_2$, after changing the variable $\kappa=k(1+it)$, is written as
\begin{subequations}\label{Square_norm_L2L2*}
	\begin{align}\label{Square_norm_L2}
		\Av{\|\hat{\mathcal{L}}\|^2}_{\mathcal{L}_2}=
		|\Xi|^2\frac{ik}{\pi}\int\limits_{0}^{L}dX\, W(X)e^{ikX}
		\int\limits_0^{\infty}dt\frac{e^{-kX t}}{\left|\zeta_0+\sqrt{1-(1+it)^2}\right|^2}\ .
	\end{align}
It must be summed up with the contribution of  segment $\mathrm{\mathcal{L}^*_2}$, which, after the change of variable $\kappa=k(1-it)$, takes the form
	\begin{align}\label{Square_norm_L2*}
		\Av{\|\hat{\mathcal{L}}\|^2}_{\mathcal{L}^*_2}=
		|\Xi|^2\frac{(-ik)}{\pi}\int\limits_{0}^{L}dX\, W(X)e^{-ikX}
		\int\limits_0^{\infty}dt\frac{e^{-kX t}}{\left|\zeta_0+\sqrt{1-(1-it)^2}\right|^2}\ .
	\end{align}
\end{subequations}
Regardless of whether parameter $kX$ in the exponents of formulas \eqref{Square_norm_L2L2*} is large or small as compared to unity, both of the integrals over $t$, for any values of $kr_c$, are well approximated by the same integral
\begin{align}\label{Approx_Int_t}
	\int\limits_0^{\infty}dt\frac{e^{-kX t}}{\left|\zeta_0+\sqrt{1+t^2}\right|^2}\approx
	\begin{cases}
		kX &\text{if}\quad kr_c\ll 1\ ,\\
		\frac{1}{kX|\zeta_0|^2} &\text{if}\quad kr_c\gg 1\ .
	\end{cases}
\end{align}
Taking into account that $X\sim r_c$, for the sum of both expressions \eqref{Square_norm_L2L2*} we obtain the estimate that coincides in order of magnitude with~\eqref{Square_norm_L1><}.

The integrals over segments $\mathcal{L}_3$ and $\mathcal{L}^*_3$ in view of the Jordan's theorem vanish.  The only point that remains for us to calculate are the integrals over contours  $\mathcal{L}_4$ and $\mathcal{L}^*_4$, i.\,e., the contribution of poles at points $k_{spp}$ and $k^*_{spp}$. The pole at point $k_{spp}$ manifests itself when calculating the term with $\widetilde{W}_+(\kappa)$. The pole contribution to the integral is calculated in a standard way and turns out to be
\begin{align}\label{Norm_pole_k-spp}
	\Av{\|\hat{\mathcal{L}}\|^2}^{(pole)}_{k_{spp}}=
	i|\Xi|^2\frac{k^2\zeta_0}{\zeta'_0 k_{spp}}
	\int\limits_{0}^{L}dX\, W(X)e^{ik_{spp}X}\ .
\end{align}
The contribution from pole $k^*_{spp}$ is complex conjugate to \eqref{Norm_pole_k-spp}, and the total pole contribution to \eqref{L_norm-def-2} takes the following form,
\begin{align}\label{Norm_pole_fin}
	\Av{\|\hat{\mathcal{L}}\|^2}^{(pole)}= &
	2|\Xi|^2\frac{k^2}{\zeta'_0} \Re \Bigg[\frac{i\zeta_0}{k_{spp}}
	\int\limits_{0}^{L}dX\, W(X)e^{ik_{spp}X}\Bigg]\approx
\notag\\
   \approx &
	2|\Xi|^2\frac{k^2}{\zeta'_0} \Re \Bigg[\frac{i\zeta_0}{k_{spp}}r_c
	\int\limits_0^{\infty}d\tau\,W(r_c\tau)e^{i(k_{spp}r_c)\tau}\Bigg]\ .
\end{align}
Under condition \eqref{Low_dissip}, the order estimate of this part of the norm squared is as follows:
\begin{align}\label{Norm_estim}
	\left|\Av{\|\hat{\mathcal{L}}\|^2}^{(pole)}\right|\sim
	\begin{cases}
		|\Xi|^2\frac{|\zeta''_0|}{\zeta'_0} kr_c &  \quad\text{if} \quad kr_c\ll 1 \ ,\\[6pt]
		|\Xi|^2 \frac{|\zeta''_0|}{\zeta'_0} &  \quad\text{if} \quad kr_c\gg 1\ .
	\end{cases}
\end{align}
Both of the limiting cases can be combined in one formula,
\begin{align}\label{Norm_estim-gen}
	\left|\Av{\|\hat{\mathcal{L}}\|^2}^{(pole)}\right|\sim
	|\Xi|^2\frac{|\zeta''_0|}{\zeta'_0}\frac{kr_c}{kr_c+1}\ ,
\end{align}
and we will consider this estimate as the final one, taking into account that the contribution of the integrals over the contours $\mathcal{L}_1$ and $\mathcal{L}_2$ is estimated by the relationship \eqref{Square_norm_L1><}, and $|\zeta_0|$,  in order of magnitude, can be set equal to unity~\cite{Palik98}.
%
\section{Binary correlators of random fields $\eta(x)$ and $\widetilde{\xi}_\pm(x)$}
\label{Bin_corr-details}
%
\subsection{``Forward" scattering}
\label{Forward_scatt}
%
Let us, first, consider a simple correlator of field  $\eta(x)$,
\begin{equation}\label{eta(x)eta(y)}
	\Av{\eta(x)\eta(y)}=\frac{1}{4 {\varkappa_{s}'}^2}
	\int\limits_{x-l}^{x+l} \frac{d x'}{2 l}\int\limits_{y-l}^{y+l} \frac{d y'}{2 l} \Av{\big[V_1(x')+V_2(x')\big]\big[V_1(y')+V_2(y')\big]}\ .
\end{equation}
By the term ``simple'' we mean that both of the random functions subject to averaging have no complex conjugation sign.

For simplicity, we will assume that random function $\zeta(x)$ is distributed according to Gaussian law, which immediately results in a zero for the cross correlator of  potentials $V_1$ and $V_2$.  Based on representation \eqref{Basic_corrs(x)} for binary impedance correlators, which, certainly, assumes that the correlation radius is small compared to the length of the interval $\mathbb{L}$,  the correlator of smoothed potential $V_1$ is represented as
\begin{align}\label{<eta_1eta_1>}
	\Av{\eta(x)\eta(y)}_{11} =& \bm{\Xi}^2\frac{k^4}{{\varkappa_{s}'}^2}\zeta_0^2
	\int\limits_{x-l}^{x+l} \frac{d x'}{2 l}\int\limits_{y-l}^{y+l} \frac{d y'}{2l} W(x'-y')=
\notag\\
    =& \bm{\Xi}^2\frac{k^4}{{\varkappa_{s}'}^2}\zeta_0^2
	\int\limits_{-\infty}^{\infty}\frac{dq}{2\pi}\widetilde{W}(q)
	\frac{\sin^2(ql)}{(ql)^2}e^{iq(x-y)}\ .
\end{align}
Choosing the sub-averaging length $l$ so that, in addition to the conditions \eqref{lambda<l<L_sc-new},  inequality $l\gg r_c$ also holds true (which is quite natural under  condition  \eqref{r_c<<L}), we take Fourier-transform $\widetilde{W}(q)$ out of the integral at the point $q=0$. Then we get
\begin{equation}\label{<11>->F_l}
	\av{\eta(x)\eta(y)}_{11}\approx \bm{\Xi}^2\frac{k^4}{{\varkappa_{s}'}^2}\widetilde{W}(0)\zeta_0^2
	F_l(x-y)\ ,
\end{equation}
where
\begin{equation}\label{F_l(x)}
	F_l(x)=\int\limits_{-\infty}^{\infty}\frac{dq}{2\pi}
	\frac{\sin^2(ql)}{(ql)^2}e^{iqx}=
	\frac{1}{2l}\left(1-\frac{|x|}{2l}\right)\Theta(2l-|x|)
\end{equation}
is the function sharp in the sense of the distribution theory, $F_l(x) \xrightarrow[l\to 0]{} \delta(x)$. The coefficient before this function in Eq.~\eqref{<11>->F_l} (see Refs.~\cite{bib:MakTar98,bib:MakTar01}) is equal to the reciprocal of the length of ``forward'' scattering  due to potential  $V_1$,
\begin{equation}\label{1/Lf-1}
	\frac{1}{L_f^{(1)}}=\bm{\Xi}^2\frac{k^4}{{\varkappa_{s}'}^2}\zeta_0^2 \widetilde{W}(0)
	\sim|\bm{\Xi}|^2k^2 r_c\ .
\end{equation}
The ``forward'' scattering length  due to potential  $V_2(x)$  is calculated in a similar way.  Keeping only this potential in Eq.~\eqref{eta(x)eta(y)}, we obtain
\begin{align}\label{<eta_1eta_2>}
	\av{\eta(x)\eta(y)}_{22} &={\bm{\Xi}^4}\frac{k^4}{2{\varkappa_{s}'}^2}
	\int\limits_{x-l}^{x+l} \frac{d x'}{2 l}\int\limits_{y-l}^{y+l} \frac{d y'}{2l} W^2(x'-y')=
	\notag\\
	&= \bm{\Xi}^4\frac{k^4}{2{\varkappa_{s}'}^2}
	\iint\limits_{-\infty}^{\ \ \infty}\frac{dqdq'}{(2\pi)^2}
	\widetilde{W}(q)\widetilde{W}^*(q')\left[\frac{\sin(q-q')l}{(q-q')l}\right]^2
	e^{i(q-q')(x-y)} \approx
	\notag\\
	&\approx \bm{\Xi}^4\frac{k^4}{2{\varkappa_{s}'}^2}
	\int\limits_{-\infty}^{\infty}\frac{dq}{2\pi}|\widetilde{W}(q)|^2\cdot
	F_l(x-y)\ ,
\end{align}
wherefrom the expression for the inverse  ``forward'' scattering length due to potential  $V_2(x)$ is obtained by analogy with \eqref{<11>->F_l},
\begin{equation}\label{1/Lf-2}
	\frac{1}{L_f^{(2)}}=\bm{\Xi}^4\frac{k^4}{2{\varkappa_{s}'}^2}
	\int\limits_{-\infty}^{\infty}\frac{dq}{2\pi}|\widetilde{W}(q)|^2\ .
\end{equation}
The ratio of reciprocal lengths \eqref{1/Lf-1} and \eqref{1/Lf-2} is asymptotically estimated as
\begin{equation}\label{Lf(1)/Lf(2)}
	\frac{L_f^{(1)}}{L_f^{(2)}}\sim \frac{|\bm{\Xi}|^2}{|\zeta_0|^2}\ ,
\end{equation}
which can take either small as compared to unity value or, in general, to reach the values of order unity.
%
\subsection{``Backward" scattering }
\label{Backward_scatt}
%
Consider the scattering produced by random fields \eqref{xi-def-new}. As in the previous section, let us start with potential $V_1(x)$. The correlator of the ``bare'' fields  $\xi^{(1)}_\pm(x)$ is calculated to
\begin{align}\label{<xi1_pm(x)xi1_pm(y)>}
	\Av{\xi^{(1)}_\pm(x)\xi^{(1)}_\pm(y)}=
	\bm{\Xi}^2\frac{k^4}{{\varkappa_{s}'}^2}\zeta_0^2
	\int\limits_{-\infty}^{\infty} & \frac{dq}{2\pi}\widetilde{W}(q)
	e^{i(q\pm 2\varkappa'_{s})(x-y)} e^{\mp 4i\varkappa'_{s}y} \times
\notag\\
   & \times\
	\frac{\sin[(q\pm 2\varkappa'_{s})l]}{(q\pm 2\varkappa'_{s})l}
	\cdot\frac{\sin[(q\mp 2\varkappa'_{s})l]}{(q\mp 2\varkappa'_{s})l}\ .
\end{align}
Due to the mismatch of the maxima of sharp factors in the integrand, the integration here will result in an additional small parameter in comparison with  \eqref{<eta_1eta_1>}, and therefore we will further neglect such correlators. The correlator of bare field $\xi^{(2)}_\pm(x)$ also contains an excessive small parameter and will be further omitted.

Consider now the correlation of the field \eqref{xi-def-new}  and its complex conjugate counterpart. The correlator of random fields associated with potential $V_1(x)$ is written as
\begin{align}\label{<xi1_pm(x)xi1_pm(y)*>}
	\Av{\xi^{(1)}_\pm(x)\big[\xi^{(1)}_\pm(y)\big]^*}=
	|\bm{\Xi}|^2|\frac{k^4}{{\varkappa_{s}'}^2}\zeta_0|^2
    \int\limits_{-\infty}^{\infty}\frac{dq}{2\pi} & \widetilde{W}(q)\,
	e^{i(q\pm 2\varkappa'_{s})x-i(q\pm 2\varkappa'_{s})y}\ \times
\notag\\
  & \qquad\times
	\left[\frac{\sin[(q\pm 2\varkappa'_{s})l]}{(q\pm 2\varkappa'_{s})l}\right]^2\ .
\end{align}
The last factor in the integrand of \eqref{<xi1_pm(x)xi1_pm(y)*>}  has maxima at points
\begin{equation}\label{q_s}
	q_s^{(\pm)}=\pm 2\varkappa'_{s}\ .
\end{equation}
The regions near these points, where the indicated factor is not parametrically small, is estimated by radius $\sim 1/l$. Assuming that inequality $l\gg r_c$ holds,  Eq.~\eqref{<xi1_pm(x)xi1_pm(y)*>} can be written in the following form,
\begin{align}\label{<xi1_pm(x)xi1_pm(y)*>_appr}
	\Av{\xi^{(1)}_\pm(x)\big[\xi^{(1)}_\pm(y)\big]^*}\approx
	|\bm{\Xi}|^2\frac{k^4}{{\varkappa_{s}'}^2}|\zeta_0|^2
	\widetilde{W}(2\varkappa'_{s})
	\int\limits_{-\infty}^{\infty}\frac{dq}{2\pi}
	e^{i(q\pm 2\varkappa'_{s})(x-y)}
	\left[\frac{\sin[(q\pm 2\varkappa_{s})l]}{(q\pm 2\varkappa_{s})l}\right]^2\ .
\end{align}
This formula is similar to expression \eqref{<11>->F_l}, so we get
\begin{equation}\label{<xi1_pm(x)xi1_pm(y)*>->F_l}
	\Av{\xi^{(1)}_\pm(x)\big[\xi^{(1)}_\pm(y)\big]^*}_{11}\approx |\bm{\Xi}|^2\frac{k^4}{{\varkappa_{s}'}^2}|\zeta_0|^2
	\widetilde{W}(2\varkappa'_{s})\cdot F_l(x-y)\ .
\end{equation}
The factor before prelimit $\delta$-function $F_l(x-y)$ is equal to the reciprocal of the scattering length due to  potential  $V_1(x)$ part carrying the momentum close to~$\pm 2\varkappa'_{s}$,
\begin{equation}\label{1/Lb-1}
	\frac{1}{L_b^{(1)}}=|\bm{\Xi}|^2\frac{k^4}{{\varkappa_{s}'}^2}|\zeta_0|^2 \, \widetilde{W}(2\varkappa'_{s})\ .
\end{equation}
Consider now the correlator  $\Av{\xi^{(2)}_\pm(x)\big[\xi^{(2)}_\pm(y)\big]^*}$ of conjugate random fields \eqref{xi-def-new} related to potential~$V_2(x)$. The statistics of  random function $\zeta(x)$ will be considered Gaussian, as before.  This correlator is calculated as
\begin{align}\label{<xi2_pm(x)xi2_pm(y)*>}
   &	\Av{\xi^{(2)}_\pm(x)\big[\xi^{(2)}_\pm(y)\big]^*}=
\notag\\
   & =\frac{k^4}{4{\varkappa_{s}'}^2}
	\int\limits_{x-l}^{x+l} \frac{d x'}{2 l} \int\limits_{y-l}^{y+l} \frac{dy'}{2l}
	e^{\pm 2i\varkappa'_{s}(x'-y')}
	\Av{\big[\zeta^2(x')-\bm{\Xi}^2\big]
		\big[\zeta^2(y')-\bm{\Xi}^2\big]^*}=
	\notag\\
	&= \frac{k^4}{2{\varkappa_{s}'}^2} |\bm{\Xi}|^4
	\int\limits_{x-l}^{x+l} \frac{d x'}{2 l} \int\limits_{y-l}^{y+l} \frac{dy'}{2l}
	e^{\pm 2i\varkappa'_{s}(x'-y')} W^2(x'-y')=
	\notag\\
	&= \frac{k^4}{2{\varkappa_{s}'}^2} |\bm{\Xi}|^4
	\iint\limits_{-\infty}^{\ \ \infty}\frac{dqdq'}{(2\pi)^2}
	e^{i(q-q'\pm 2\varkappa'_{s})(x-y)}\widetilde{W}(q)\widetilde{W}^*(q')
	\left[\frac{\sin\big[(q-q'\pm 2\varkappa'_{s})l\big]}
	{(q-q'\pm 2\varkappa'_{s})l}\right]^2\ .
\end{align}
Using the sharpness of the factor in square brackets in comparison with function $\widetilde{W}^*(q')$  the latter can be taken out of the integral over $q'$ at points $q'=q\pm 2\varkappa'_{s}$ giving the result
\begin{align}\label{<xi2_pm(x)xi2_pm(y)*>-appr}
	\Av{\xi^{(2)}_\pm(x)\big[\xi^{(2)}_\pm(y)\big]^*}\approx
	\frac{k^4}{2{\varkappa_{s}'}^2} |\bm{\Xi}|^4
	\int\limits_{-\infty}^{\infty}\frac{dq}{2\pi}
	\widetilde{W}(q)\widetilde{W}^*(q\pm 2\varkappa'_{s})\cdot F_l(x-y)\ .
\end{align}
Comparing this formula with \eqref{<xi1_pm(x)xi1_pm(y)*>->F_l} we obtain the contribution of the potential  $V_2(x)$ to the damping rate associated with the backscattering,
\begin{equation}\label{1/Lb-2}
	\frac{1}{L_b^{(2)}}=\frac{k^4}{2{\varkappa_{s}'}^2}|\bm{\Xi}|^4 \int\limits_{-\infty}^{\infty}\frac{dq}{2\pi}
	\widetilde{W}(q)\widetilde{W}^*(q\pm 2\varkappa'_{s})\ .
\end{equation}
%
\subsection{``Mixed" correlator}
\label{Mixed_scatt}
%
For correlator $\av{\xi_\pm(x)\eta(y)}$ which, to be specific, we call ``mixed'' correlator, the fundamental inability to remove rapidly oscillating phase factor $e^{\pm 2i\varkappa'_{s} x'}$ appearing in Eq.~\eqref{xi-def-new} is extremely important. For potential $V_1(x)$ this correlator is calculated to
\begin{align}\label{Mixed_corr}
	\av{\xi_\pm(x)\eta(y)}_{11}&=\left(\frac{k^2}{\varkappa'_{s}}\right)^2\big(\zeta_0\bm{\Xi}\big)^2
	\int\limits_{x-l}^{x+l}\frac{d x'}{2 l}e^{\pm 2i\varkappa'_{s} x'}
	\int\limits_{y-l}^{y+l}\frac{d y'}{2 l}W(x'-y')=
	\notag\\
	&= \frac{k^4}{{\varkappa_{s}'}^2}\bm{\Xi}^2\zeta_0^2 e^{\pm 2i\varkappa'_{s}x}
	\int\limits_{-\infty}^{\infty}\frac{dq}{2\pi}e^{iq(x-y)}\widetilde{W}(q)
	\,\frac{\sin\big[(q\pm 2\varkappa'_{s})l\big]}{(q\pm 2\varkappa'_{s})l}\cdot
	\frac{\sin(ql)}{ql}\ .
\end{align}
The mismatch of the sharp functions maxima of in the integrand results in that correlator  \eqref{Mixed_corr} is parametrically small in comparison with correlators \eqref{<11>->F_l} and \eqref{<xi1_pm(x)xi1_pm(y)*>->F_l} in the integral sense, and therefore, we will neglect all correlators of this type assuming that random fields $\eta(x)$ and $\xi_{\pm}(x)$ are not correlated.
%
%

\end{document}